\documentclass[preprint2]{aastex}
\usepackage{graphicx}

\newcommand{\ha}{H\ensuremath{\alpha}}

\newcommand{\hii}{H{\scshape ii}}
\newcommand{\nii}{[N{\scshape ii}]}   
\newcommand{\oiii}{[O{\scshape iii}]} 

\def\degree{\mbox{$^{\circ}$}}

\slugcomment{Accepted by AJ, 27 December 2010}
\shorttitle{Properties of the HII regions of M51 and NGC 4449.}
\shortauthors{Guti\'{e}rrez \& Beckman}

\begin{document}

\title{Properties of the HII region populations of M51 and NGC 4449 \\ 
       from \ha{} images with ACS on HST.}

\author{Leonel Guti\'{e}rrez\altaffilmark{1,2,4}, John E. Beckman\altaffilmark{2,3,4}, and Valeria Buenrostro\altaffilmark{2,4,5}}
\altaffiltext{1}{Universidad Nacional Aut\'onoma de M\'exico, Instituto de Astronom\'{\i}a, Ensenada, B. C. M\'exico}
\altaffiltext{2}{Instituto de Astrof\'{\i}sica de Canarias, C/ Via L\'{a}ctea s/n, 38200 La Laguna, Tenerife, Spain}
\altaffiltext{3}{Consejo Superior de Investigaciones Cient\'{\i}ficas, Spain}
\altaffiltext{4}{Facultad de F\'{\i}sica, Universidad de La Laguna, Avda. Astrof\'{\i}sico Fco. S\'{a}nchez s/n, 38200, La Laguna, Tenerife, Spain}
\altaffiltext{5}{Universidad Nacional Aut\'onoma de M\'exico, Centro de Radioastronom\'{\i}a y Astrof\'{\i}sica, Morelia, M\'exico}

\email{leonel@astrosen.unam.mx, jeb@iac.es, v.buenrostro@crya.unam.mx}

\begin{abstract}
We have used the images from the ACS on HST in \ha{}, and in the neighboring continuum, to produce flux calibrated images of the large spiral galaxy M51, and the dwarf irregular NGC 4449. From these images we have derived the absolute luminosities in \ha{}, the areas, and the positions with respect to the galactic centers as reference points, of over 2600 \hii{} regions in M51 and over 270 \hii{} regions in NGC 4449. Using this database we have derived luminosity (L)--volume (V) relations for the regions in the two galaxies, showing that within the error limits these obey the equation $L \sim V^{2/3}$, which differs from the linear relation expected for regions of constant uniform electron density. We discuss briefly possible models which would give rise to this behavior, notably models with strong density inhomogeneities within the regions. Plotting the luminosity functions for the two galaxies we find a break in the slope for M51 at $\log(L)$ = 38.5 dex (units in erg s$^{-1}$) for M51 in good agreement with the previous ground-based study by Rand, and above this luminosity NGC 4449 also shows a sharp decline in its luminosity function, although the number of regions is too small to plot the function well at higher luminosities. The cumulative diameter distribution for the \hii{} regions of M51 shows dual behaviour, with a break at a radius close to 100 pc, the radius of regions with the break luminosity. Here too we indicate the possible physical implications.
\end{abstract}


\section{Introduction}\label{sec:intro}

There is a well grounded tradition of studying populations of \hii{} regions in spiral and 
irregular galaxies, which goes back to the work of Hodge, Kennicutt, and their 
collaborators \citep{kennicutt80, hodge89, kennicutt89}. It has become customary to measure their \ha{} 
luminosities and their radii, and prepare systematic catalogs. As techniques have improved and 
the precision of the measurements has increased the scope of the measurements has been gradually 
extended. The methods for discriminating an \hii{} region from its diffuse surrounding emission 
have been improved, and the luminosity functions have been extended to lower limiting values. 
Although the ultimate aim should be to improve our understanding of the star formation process 
and the interaction of, notably massive, stars with the ISM, working on the improvement of the 
intrinsic data base is an objective intrinsically worth pursuing.  Considerable worthwhile ground 
based data have been reported (a representative but in no way comprehensive list of references 
is: \citet{rand92, knapen93, rozas96, rozas99, rozas00, gonzalez97, feinstein97, 
youngblood99, mackenty00, thilker02, buckalew06, hakobyan08}) and, in the case 
of M33, \citet{cardwell00} presented a list of over 9,000 detected regions 
in a single galaxy,  but it is clearly of interest to analyze data of the quality and angular 
resolution provided by HST for this purpose. However relatively little work has been presented 
of this type, because the required narrow band filters at arbitrary recession velocities are 
not included in the optics of the HST instruments. A notable previous study of M51 based on 
WFPC2 narrow band imaging was presented by \citet{scoville01}, and a parallel 
study of the same galaxy based on NICMOS observations in Pa$\alpha$ was published by \citet{alonso01}. In this article we use ACS data for the first time to examine the 
statistical properties of \hii{} regions in whole galaxies, presenting results for M51, and also 
for the dwarf irregular NGC 4449.  
In section 2 we discuss how the archival data were processed in order to prepare the \ha{} maps,  
and in section 3 we describe the critically important processes of continuum and background 
subtraction, followed by the derivation of the luminosities, areas, and positions of the 
regions. In section 4 we derive the luminosity-radius relations and the luminosity functions 
for both galaxies and the diameter distribution for M51, and in section 5 we offer a discussion 
and our conclusions.   

\section{Observations, and basic data reduction}\label{sec:data}

\subsection{Basic processing}

The images of M51 (NGC 5194) and its companion NGC 5195 were observed with the wide field camera of the ACS on HST in January 2005 within the program 10452 (PI: Steve Beckwith, Hubble Heritage Team). The observations consisted of 6 pointings through two broad band filters: 
F555W (close to standard $V$), and F814W (close to $I$), and a narrow band \ha{} filter, F658N. The exposures per pointing were 
340s, 340s, and 680s 
for the three filters, respectively. The observations used here, taken from the HST archive, were corrected, calibrated, and combined by \citet{mutchler05}. 

The images of NGC4449 were also taken from the HST archive. They were taken during November 2005 through the same filters as those used for M51, within the program GO 10585 (PI: Alessandra Aloisi). The exposures were taken at two different pointings along the axis of the galaxy, with four exposures per filter at each pointing, dithered to optimize the removal of cosmic rays and bad pixels, and to allow an improved determination of the point spread function. The total, integrated exposure times for the three filters, after combining the images using MULTIDRIZZLE \citep{koek02} were 
2400s, 2000s, and 360s respectively, in 
$V$, $I$, and \ha{}. The total area imaged on the galaxy is 340$\times$200 arcsec$^2$, with pixels of 0.05 arcsec.

The archive data were processed {\it on-the-fly}, correcting for bias and dark current, flat-fielding, and correcting for non-linear effects using the standard pipeline calibration for ACS (CALACS, \citet{hack00}). The images were also corrected for detector defects using MULTIDRIZZLE, which also corrects each image for the distortion associated with each filter, and can then be used to combine images, and take out cosmic rays. For M51, these processes were performed by \citet{mutchler05}, but for NGC 4449 they were carried out by ourselves, with the help of IRAF, which was employed to combine all the images for a given pointing and a given filter into single images before joining them in a mosaic to take in the whole galaxy. This was done by making use of the overlap zone in the pointings ($\sim$ 30 $\times$ 200 arcsec$^2$). 
The alignment errors were less than 0.3 pixels in any direction, which we derived from the differences in the centroids of the positions of the stars used for positional calibration.

Images processed with MULTIDRIZZLE contain the signal in units of electrons/sec, and from the image headers one can find the sensitivity conversion factor, PHOTFLAM, defined as the mean flux $F$ 
(units of erg cm$^{-2}$ s$^{-1}$ \AA$^{-1}$),
which yield 1 count per second in the observing mode used. Any gain differences between the ACS detectors are already accounted for by the photometric correction, so that to obtain a calibrated image in erg s$^{-1}$ cm$^{-2}$ \AA$^{-1}$ one simply multiplies by the constant factor PHOTFLAM. To convert this flux image into an HST magnitude one uses the expression 

\begin{equation}
			m = -2{.}5 \times log_{10} (F) + PHOTZPT		\label{eq:n1}
\end{equation}

\noindent where the value of PHOTZPT is the zero point on the ST magnitude scale, and is usually listed in the image header, but can also be found on the ACS website (http://www.stsci.edu/hst/acs/analysis/ zeropoints). We list in Table \ref{T:M1} the values of PHOTFLAM  and ZPT for the three filters used \citep{sirianni05}. 

\begin{table*}
\begin{center}
\begin{tabular}{c c c c c}
\hline
\hline
Filter & 
\begin{tabular}{c}$\lambda_{c}$ \\ (\AA) \\ \end{tabular}  &
\begin{tabular}{c}$\Delta\lambda$ \\ (\AA) \\ \end{tabular}  &  
\begin{tabular}{c}PHOTFLAM \\ erg{$\cdot$}cm$^{-2}${$\cdot$}s$^{-1}${$\cdot$}\AA$^{-1}$ \\ \end{tabular} &
ZPT$_{STMAG}$ \\
\hline
F555W & 5359.547 & 847.78  & 1.955929E-19 & 25.672  \\
F658N & 6584.015 & 87.48  & 1.999181E-18 & 23.148  \\
F814W & 8059.761 & 1541.56  & 7.072360E-20 & 26.776  \\
\hline
\end{tabular}
\caption{\label{T:M1} Conversion factors for sensitivity and zero points for ACS/WFC. $\lambda_c$ is the central (pivot) wavelength of the filter and $\Delta\lambda$ is the full width at half maximum (FWHM) of the filter. ZPT$_{STMAG}$ is the zero point in the magnitude scale \emph{ST}.}
\end{center}
\end{table*}

\section{The HII region \ha{} luminosities}
\subsection{Continuum and background subtractions}

\begin{figure*}
  \begin{center}
  \begin{minipage}{5.0in}
   \centering
   \plotone{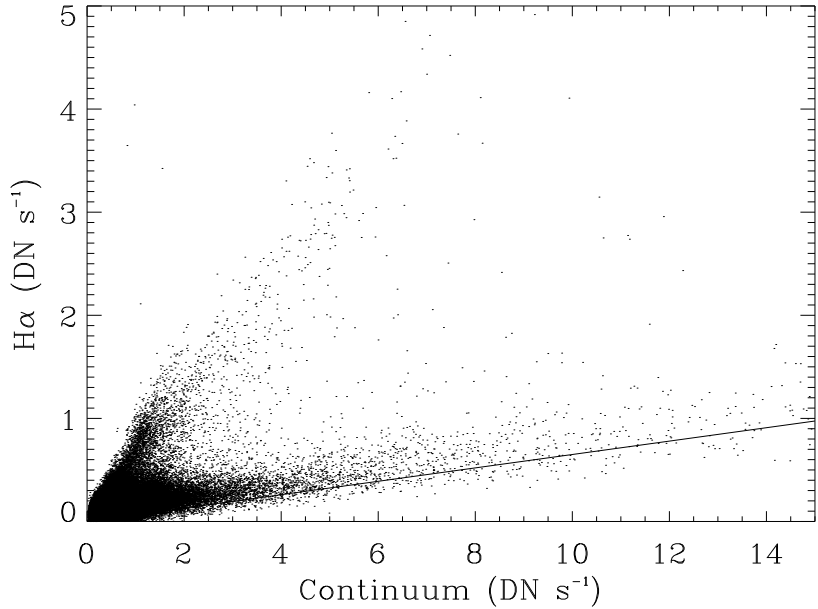} 
   \caption{Plot of the pixel intensity in the \ha{} image of the galaxy NGC 4449 against the intensity of the same pixel in the continuum image. The two images were first registered in position to a fraction of a pixel (some 0.3 pix) taking bright stars for refereence. In order not to overcrowd the plot we use a 1000 x 400 pixel portion of the galaxy to derive it.} \label{F:Continuo}
  \end{minipage}
  \end{center}
\end{figure*}

Once all the images had been aligned we could subtract the continuum contribution to the total fluxes in the \ha{} images, to isolate the flux from the emission line itself. Since we did not have a narrow band continuum image close to \ha{}, we used a weighted combination of the $V$ and $I$ images, deriving the proportions of each by minimizing the residuals of the stellar images when the subtraction of the weighted continuum from the \ha{} image had been effected. The combined image chosen for the final subtraction comprised 15\% of the $I$ image and 85\% of the $V$ image. The factor by which this combined $VI$ continuum image had to be multiplied before the subtraction in order to produce an image in \ha{} flux alone was found, using a method developed by \citet{boker99} and used by \citet{knapen04}. This consists in plotting the intensity in the \ha{} image, pixel by pixel, against the corresponding intensity in the continuum image. An example of this process is shown in Fig. \ref{F:Continuo} 
where we used the original \ha{} image and the weighted sum of the $V$ and $I$ continuum images. As most of the pixels do not contain any signal due to \ha{}, the majority of points in the plot lie on a straight line, with low scatter, whose slope gives the constant of proportionality for the continuum subtraction. The continuum-subtracted \ha{} image is then produced by subtracting the continuum image, multiplied by this constant, from the original \ha{} image.

It is possible that the signal from the F555W filter ($V$) includes some contamination
from strong emission lines, of which \oiii{}$\lambda$5007 is the strongest, because 
the filter transmission for that redshifted line is near maximum. However,
the spectroscopic measurements made by \citet{bresolin04} on 10 \hii{} regions in 
M51 show that the \oiii{} line emission ranges from 1.6\% to 56\% of the \ha{} emission 
with an average fraction of 14\%. 
On the other hand, the surface brightness in the F555W filter on the \hii{} regions is, 
on average, $\sim$3 times that in F658N, measured on our images. So, we can infer that the emission in 
\oiii{}$\lambda$5007 is on average 5\% of the continuum (with a maximum fraction of 1/6th). 
Taking into account that in the calibration process we found that the ratio of the
off-band continuum to the on-band line+continuum intensities is $\sim$0.06, we consider  
that we need not be worried about significant over-subtraction of the continuum due to
the \oiii{} emission, which would make a mean contribution of order 0.3\% (maximum 1\%) 
of the total signal.

Once the \ha{} flux map had been obtained, we went on to determine the luminosities of the individual \hii{} regions. The difficulty here lies in effecting a valid separation of the region from its surrounding background, and above all from any overlapping regions. To separate partially overlapping regions, we opted to count as two regions those whose brightness distribution between the two central maxima dipped below 2/3 of the central brightness of the fainter of the two regions, while those which did not satisfy this criterion were treated as single regions. This process was unambiguous for most cases. For the special cases of regions old enough for their centers to have been blown outwards to the extent that their brightness distributions are annular (i.e. we are looking at expanding shells), we considered such objects as single regions.

As far as the background is concerned we had to carry out the following process: define the boundary of each region, determine a value for the \ha{} background in the surrounding zone, and subtract this value from each pixel of the  defined \hii{} region before integrating the remaining flux over all the pixels to give the total flux for the region.  The \ha{} background is due to the emission of the diffuse warm ionized gas surrounding the region, and this procedure assumes that this pervades the disk, in such a way that along the line of sight to an \hii{} region we see the combined effect of emission from the region and from the diffuse gas. This assumption, and other rival assumptions, are discussed in detail in \citet{zurita04}, who reach the conclusion that it is difficult to do better than this, even though the assumption may not give a perfect description of the local morphology.

\begin{figure*}
\centering
   \begin{minipage}{5.0in}
   \plotone{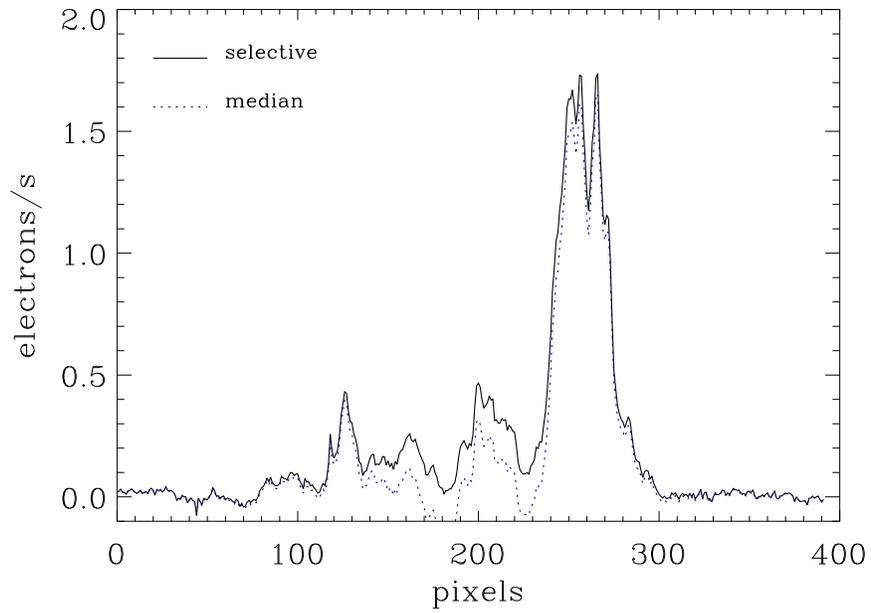}
   \caption{Intensity in electrons per second, at the 400 pixel length scan across NGC 4449 shown in Fig. \ref{F:regionconlinea}. The intensity was measured in the median filtered image (dotted line) and also in the image treated with the selective filter (solid line) as described in the text. The scans were calculated by averaging over 10 adjacent columns.} \label{F:comparamediana}
   \end{minipage}
\end{figure*}

\begin{figure*}
\centering
\begin{minipage}{5.0in}
   \plotone{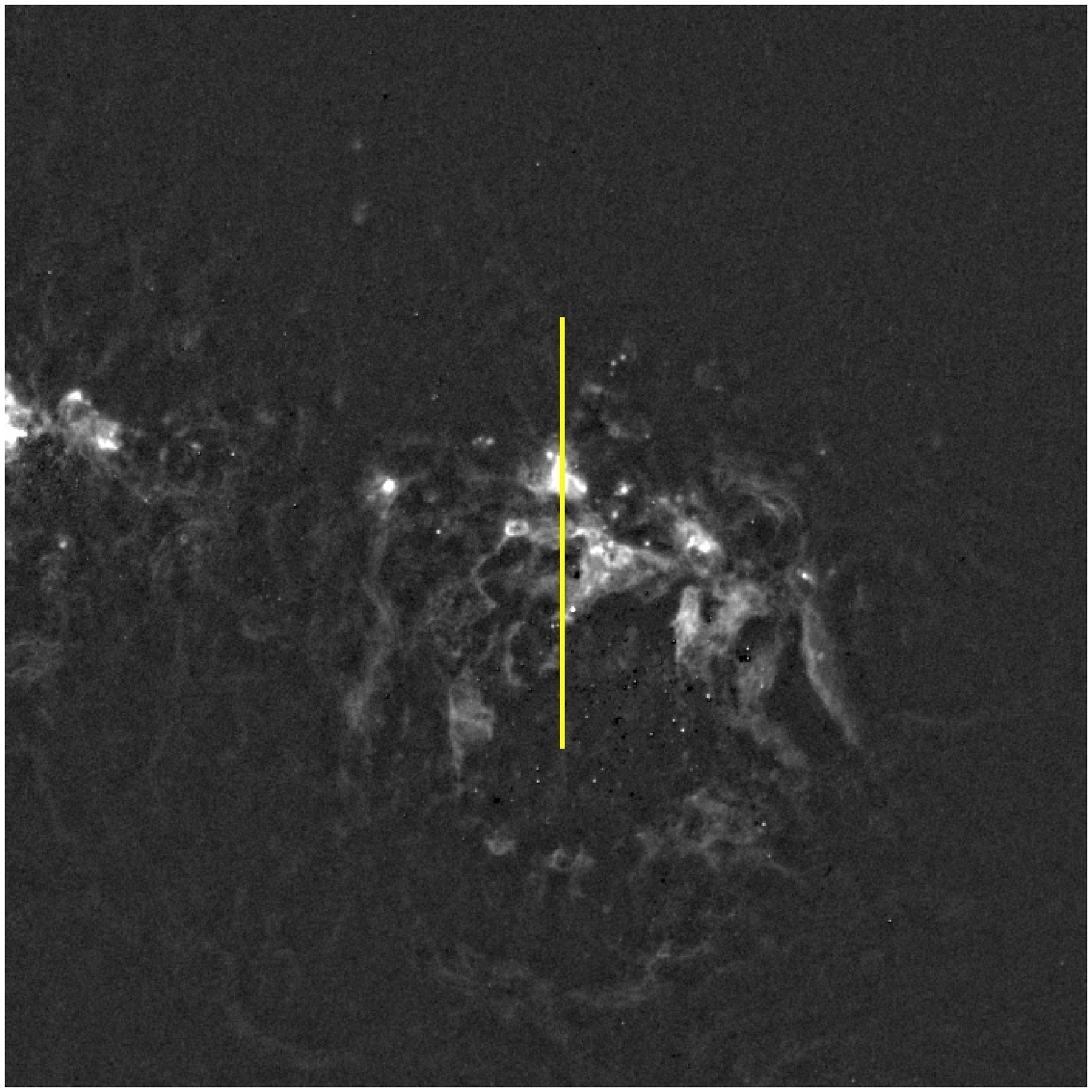}
   \caption{Ionized compled in the south-east of NGC 4449. A scan of length 400 pixels for which we calculated the intensities in Fig. \ref{F:comparamediana} is shown. The intensities are the averages of ten adjacent columns.} \label{F:regionconlinea}
\end{minipage}
\end{figure*}

For M51 we determined the background in rings around the outside of the diaphragm used to determine the \hii{} region flux. After trials, in which the ring width was varied between 2 and 5 pixels we found that the difference in estimated background, defined as the median flux value per pixel over the ring, varied by no more than 12\%, and we confined our measurements to rings of width 2 pixels, before subtracting this value from the pixels of the region, and integrating to give the total flux over each region. For NGC 4449 the task was more complex due to the presence of strong diffuse emission, which had significant gradients, and we opted to make a map of the diffuse emission and correct for it before subtracting the local backgrounds. We designed a selective filter to do this, since we had previously determined that a median filter tended to yield backgrounds which were too strong, and whose subtraction led to underestimates of the \hii{} region fluxes. We used boxes of 101 pixels on the side, and to estimate the background within a box we first calculated the median value, then discarded the pixels in the upper and lower 5\% flux ranges, and then tested to see whether the fluxes in all the pixels below the median were within a 3$\sigma$ limit (we had previously determined the standard deviation $\sigma$ in a zone which did not contain \hii{} regions). If this test was not satisfied, the program reduced the percentile progressively until the pixel fluxes did satisfy the criterion, down to a limiting 10 percentile. Finally the program computed an image which gave an excellent fit to the diffuse background, and we could make the appropriate subtraction systematically. In Fig. \ref{F:comparamediana} %
we show a scan across a typical zone of the galaxy with one major \hii{} region and several minor regions (see Fig. \ref{F:regionconlinea}), to which we had applied a median filter correction (dotted line), and a correction using the selective filter (solid line), which demonstrates the value of our method. So our final procedure for this galaxy was to use our selective filter to filter out the large scale features corresponding to the diffuse background, and then subtract off the local background flux from each region, as described briefly below.

\subsection{The luminosities, areas, and positions of the population of HII regions}

The flux within an \hii{} region is derived by integrating the intensity over all those contiguous pixels making a continuous object, whose intensity is greater than three times the rms scatter of the local background. From this integrated value we then subtract off the product of the number of pixels so sampled by the flux value per pixel of the local background. These fluxes were converted from electrons s$^{-1}$ to erg s$^{-1}$ cm$^{-2}$ using PHOTFLAM as described above, and using the best estimated distances to M51 or to NGC 4449 we could convert these values to absolute luminosity in erg s$^{-1}$ for a given \hii{} region. After considering and comparing the values of the distance to M51 by a number of authors, we selected the distance to M51 given by \citet{feldmeier97} of 8.39 Mpc, based on the luminosity function of its planetary nebulae, as the most reliable. This measurement, together with that of \citet{tonry01} using surface brightness fluctuations on NGC 5195, with which it agrees within the limits of their errors, gives the lowest scatter and systematic errors of the techniques published to date for this observation. At this distance, the scale of the image is equivalent to 2.03 pc pixel$^{-1}$. For NGC 4449 we have used a distance of 3.82 Mpc from \citet{annibali08} who used the tip of the red giant branch. This value is the most accurate one published so far, and is also in fair agreement with the results of \citet{karachentsev03}, who used the same technique, but gave wider error bars. At this distance the image scale is 0.93 pc pixel$^{-1}$. 

From the flux measurements we could also derive an area $A$ for each \hii{} region, multiplying the area per pixel by the number of pixels counted as part of the region as defined above, and we could then measure an {\it equivalent radius}, $R_{eq}$, defined by $R_{eq} = \sqrt{ (A/\pi)}$. The position of each region was determined by the centroid of the region which was found in rectangular coordinate directions, measured from the galactic nucleus, by taking the flux weighted sum of the coordinate distances of the pixels from the nucleus along each coordinate direction divided by the integrated flux along that direction. For regions in the form of shells this centroid did not, in general, lie within a luminous part of the region, though for the majority of regions it lay close to the visually inspected center.

The luminosities in \ha{} for the \hii{} regions had to be corrected for the presence of the  \nii{} doublet, $\lambda\lambda$6548,6584 \AA, as the narrow band filter F658N is wide enough to  take in both of these emission lines. This cannot be done region by region, but using the detailed measurements for M51 by \citet{bresolin04} of the strengths of the two doublet lines in ten regions at distances ranging from 0.19 to 1.04 of the isophotal radius, we find that to use a constant factor, independent of radius, of 
\nii{}$\lambda\lambda$6548,6584 equal to 0.33 ($\pm{}$0.06) found as the mean value of Bresolin et al.\'\,s regions (in agreement with the value used by \citet{hill97}), gives an entirely satisfactory procedure for M51, given the very moderate metallicity gradient found in the disk of that galaxy. For the wavelengths of the two \nii{} doublet lines, and of \ha{}, redshifted for M51\'\,s recession velocity, the differences in transmission across the F658N filter are less than 1\%, and could be neglected. So, the flux measured in F658N was multiplied by 0.67 to give the \ha{} flux. For the redshift of NGC 4449, however, the transmissions of the filter are 0.730, 0.923, and 0.929, at the wavelengths of the three lines, in ascending wavelength order, respectively, and appropriate corrections had to be applied. From \citet{osterbrock89} we know that  \nii{}$\lambda$6548/\nii{}$\lambda$6584 = 0.33 and for NGC 4449 we find from \citet{kennicutt92} that
\nii{}$\lambda$6584/\ha{} =0.12. Making the transmission corrections we deduce that the flux measured in F658N should be multiplied by 0.862 to give the \ha{} flux. This single factor is deemed adequate given that the metallicity measured close to the nucleus by \citet{boker01} is in agreement with the metallicity in the disk measured by \citet{lisenfeld98} within the limits of their mutual errors.

We have described in this section the derivations of the \ha{} luminosities, equivalent radii, and positions of the \hii{} regions measured in M51 and NGC 4449. These are listed in Tables 3a and 4a
in the on-line version of this article in the journal, and the data for the 50 most luminous regions in each galaxy is listed here in Tables \ref{T:RegionsM51} and \ref{T:RegionsNGC}

\section{Basic physical properties of the populations of HII regions in the two galaxies, as derived from their \ha{} emission}

\subsection{The luminosity-volume relations}

\begin{figure*}
\centering
\begin{minipage}{5.0in}
   \plotone{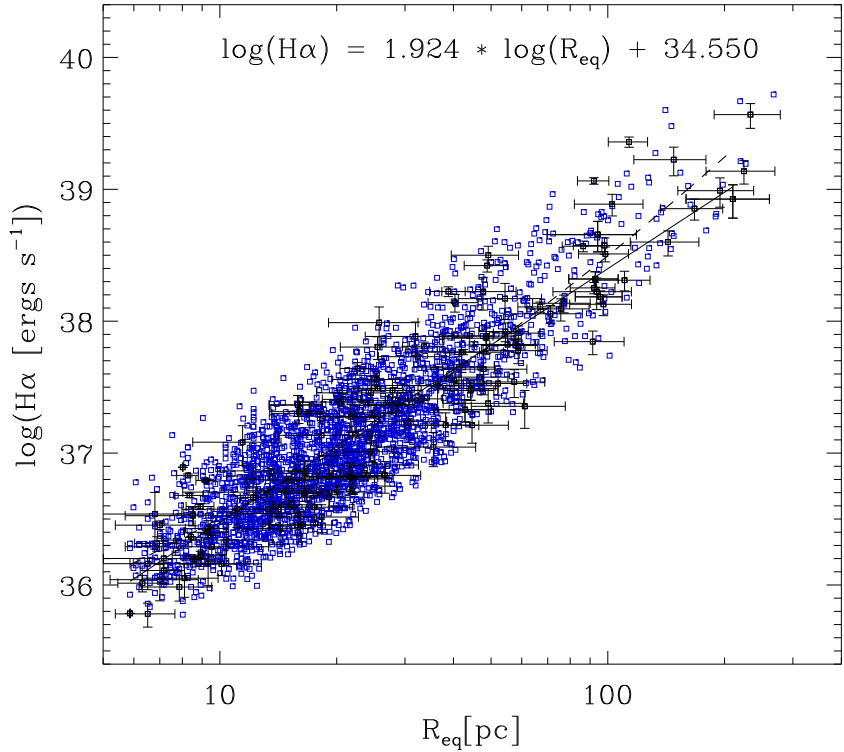}
   \caption{ Relation between the logarithm of the \ha{} luminosity in erg s$^{-1}$ and $R_{eq}$ in parsec on logarithmic scales, for the measured \hii{} regions of M51. The solid line shows the best linear fit to the distribution $log(L_{H\alpha}) =  1{.}924{\times}log(R_{eq}) + 34{.}550$ and the dotted line the even better second order fit $log(L_{H\alpha}) = 35{.}35 +  0{.}78{\times}log(R_{eq}) +  0{.}40\times(log(R_{eq}))^2$. Error bars are plotted for a random selection of 150 of the \hii{} regions (including more error bars would lead to a diagram too confused for ready interpretation). In general the uncertainties in determining the areas of the regions are, as seen, greater than those found in the measurements of their luminosities, because the major fraction of the luminosity is concentrated in the center of each region.} \label{F:lum_radio}
\end{minipage}
\end{figure*}

In  Fig. \ref{F:lum_radio} %
we have plotted the distribution of the radii of the \hii{} regions in M51 against their \ha{} luminosities, using the data tabulated in Table
3a (on-line version of the Journal).
In this log-log plot we can see that there is a reasonable first order straight line fit to this large collection of data points, which takes the form:

\begin{equation}
			log (L_{\ha{}}) = 1.92log (R_{eq}) +34.55	\label{Eq:n2}
\end{equation}

\noindent where $R_{eq}$ is the equivalent radius of the \hii{} region defined above. A more accurate fit to the distribution is given by the quadratic expression

\begin{equation}
	log (L_{\ha{}}) = 35.35 + 0.78 log (R_{eq}) + 0.40(\log(R_{eq}))^2   \label{Eq:n3}
\end{equation}

We know that if \hii{} regions were spherical and with uniform density, the fit should be linear and have slope 3, in accordance with  ``Case B'' of Osterbrock (1989), \citep[or indeed with the classical model of][]{stromgren39}, as given by the relation:

\begin{equation}
	L(\ha{}) = \frac{1}{2{.}2} \left\{\frac{4\pi{}N_{H}^2\alpha_{B}hc}{3~\lambda_{\ha{}}}\right\} R^3	\label{Eq:n4}
\end{equation}

\noindent where $N_{H}$ is the density of the hydrogen in the region in atoms cm$^{-3}$, which we assume to be equal to the electron density $n_e$,  $\alpha_B$ is the recombination coefficient for all the excited levels of hydrogen, and $\lambda_{\ha{}}$ is 6563\AA. A  part of the discrepancy between prediction and observation may be due to the fact that a fraction of the luminosity of the smaller regions is buried in the noise, so that the slope of the relation should (and does) increase somewhat with increasing luminosity. However the basic difference cannot be explained in this way, and we will return to this point in a later section of the paper. If we do take the expression in Eq. \ref{Eq:n4} as valid, we can calculate an equivalent mean electron density, which varies between 1 and 30 cm$^{-3}$ with a mean value $\left<n_e\right>$ close to 6 cm$^{-3}$. 

\begin{figure*}
\centering
\begin{minipage}{5.0in}
   \plotone{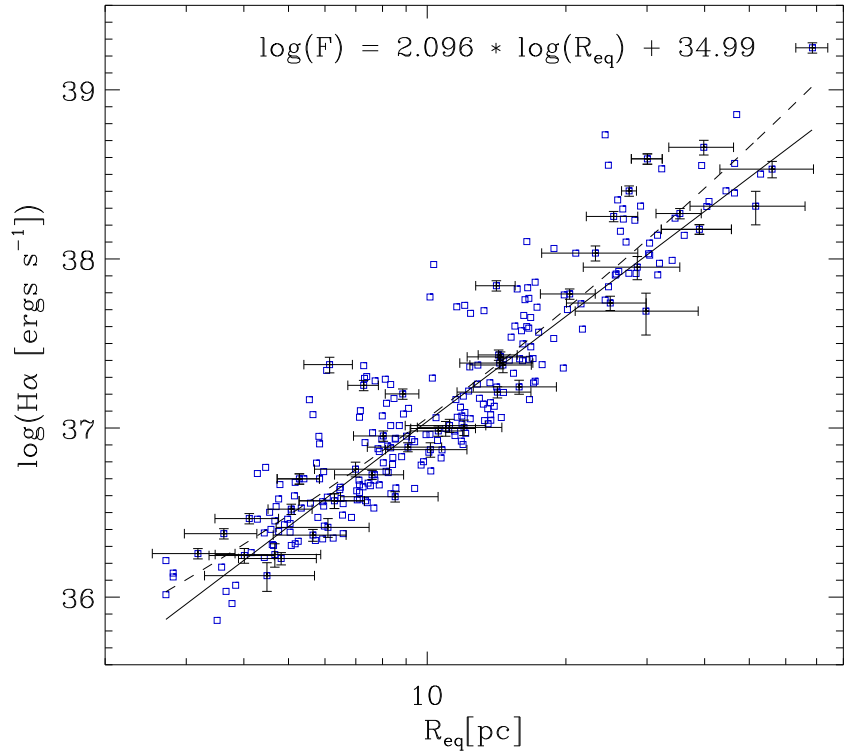}
   \caption{ Relation between the logarithm of the \ha{} lumnosity in erg s$^{-1}$ and  $R_{eq}$ in parsec, also logarithmic, for the measured \hii{} regions in NGC 4449. The solid line shows the best linear fit to the distribution $log(L_{H\alpha}) =  2{.}096{\times}log(R_{eq}) + 34{.}99$, and the dotted line shows the even better second order fit $log(L_{H\alpha}) = 35{.}41 +  1{.}26{\times}log(R_{eq}) +  0{.}38\times(log(R_{eq}))^2$. As in Fig. 4, error bars
are plotted for a random selection of 50 \hii{} regions.} \label{F:lum_radio_n4449}
\end{minipage}
\end{figure*}

In Fig. \ref{F:lum_radio_n4449}
 we show the plot of equivalent radius against \ha{} luminosity for the population of the \hii{} regions of NGC 4449. The best linear fit to the points is given by:

\begin{equation}
\log{(L_{\ha{}})} = 2{.}09 \log{R_{eq}} + 34{.}99     \label{Eq:n5}
\end{equation}

\noindent and a more accurate fit is given by the quadratic expression:

\begin{equation}
log (L(\ha{})) = 35.41 + 1.26 \log(R_{eq}) + 0.38(\log(R_{eq}))^2 \label{Eq:n6}
\end{equation}

These expressions are very similar to those for M51, and in particular for both galaxies the luminosities in \ha{} vary as the square of the radii of the regions (see Eqs. \ref{Eq:n2} and \ref{Eq:n5}) rather than as the cube, which would be predicted from standard models of \hii{} regions.

\subsection{The luminosity functions of the HII 
regions in both galaxies}

\begin{figure*}
\centering
\begin{minipage}{5.0in}
   \plotone{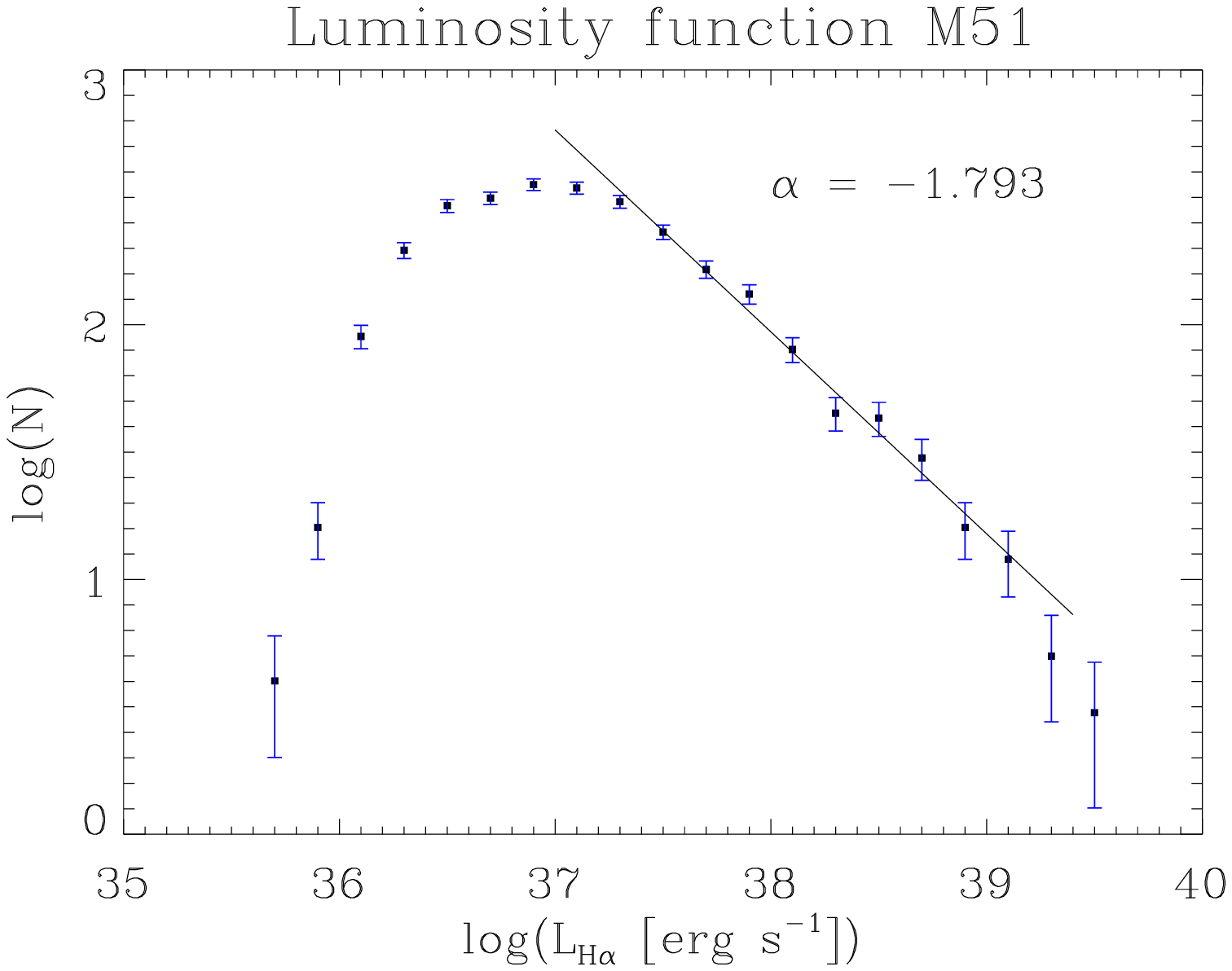}
   \caption{Luminosity function of the \hii{} regions of M51. We plotted 
   the number $N$ of observed regions with luminosity between $L$ and $\Delta{L}$ 
   (erg s$^{-1}$ on a logarithmic scale), with  $\Delta{L}$ = 0.2 dex.
} \label{F:lum_func}
\end{minipage}
\end{figure*}

In Fig. \ref{F:lum_func} 
we show the luminosity function of the \hii{} regions in M51, plotted as the log of the number of regions in bins of width 0.2 dex in luminosity against the log of the luminosity itself. We have fitted a power law fit to the declining portion of the function towards high luminosities, using the equation:

\begin{equation}
N(L){\rm d}L=AL^{\alpha}{\rm d}L,    \label{Eq:n7}
\end{equation}

\noindent where $N(L){\rm d}L$ is the number of \hii{} regions with luminosities in the interval between $L$ and $L+dL$ (see Kennicutt et al. 1989). The slope of this straight line fit in the log-log diagram is -0.793, which implies that $\alpha$ = -1.793, because the luminosity function is plotted with logarithmic bins, whereas the slope $\alpha$ is defined via the differential function in Eq. \ref{Eq:n7} with linear bins. The interval over which we have determined $\alpha$ in Fig. \ref{F:lum_func} is from $\log(L)$ = 37.2, to $\log(L)$ = 39.2 , i.e. a hundredfold interval in luminosity. Below the lower limit the function is incomplete for S:N considerations, and above the upper limit the function becomes uncertain due to the paucity of the statistics.

We next effect a separation of the \hii{} regions in the arms from those in the interarm zones. To do this we made a mask using the I image, and after applying a median spatial filter, using a 10$\times$10 arcsec box, to reduce local effects of dust structure, we then binned into 10$\times$10 pixel bins to reduce the computational load, and finally,  using a PA of 26$\degree$~ and an inclination of 42$\degree$~ \citep{bersier94}, we fitted an artificial exponential disk to the galaxy in such a way that 50\% of the light from the galaxy disk lay above this disk, and 50\% below it. To assign regions to the arms or interarm zones we set a mask such that the part of the galaxy above this discriminating exponential was set to unity (the arms) and the rest of the galaxy set to zero (the interarm zones). As the galaxy is deformed by its interaction with NGC 5195 we had to make manual corrections to the mask for those (relatively small) outer parts of the galaxy where the single exponential fit was clearly inadequate. To illustrate our results we show, in Fig. \ref{F:lum_func1},
the distribution of the \hii{} regions in the two spiral arms, plotted in polar coordinates on the deprojected image of the disk. In Fig. \ref{F:lum_func2}
we show the luminosity functions of the arm \hii{} regions, the interarm regions and, for comparison, of the full population, which is the same as the plot in Fig. \ref{F:lum_func}. In total we measured luminosities for 1966 regions in the arms, or within the crowded zone within 1 kpc of the galaxy center, and 693 regions in the interarm zones. The integrated luminosity of all the regions in the arms and in the central 1 kpc is 1.25$\times$10$^{41}$ erg s$^{-1}$ (90\% of the total) and the integrated luminosity of the interarm \hii{} regions is 1.37$\times$10$^{40}$ erg s$^{-1}$ (some 10\% of the total). If we exclude the central 1 kpc the integrated luminosity of the regions in the arms is 1.18$\times$10$^{41}$ erg s$^{-1}$, some 85\% of the total. 

\begin{figure*}
\centering
\begin{minipage}{5.0in}
   \plotone{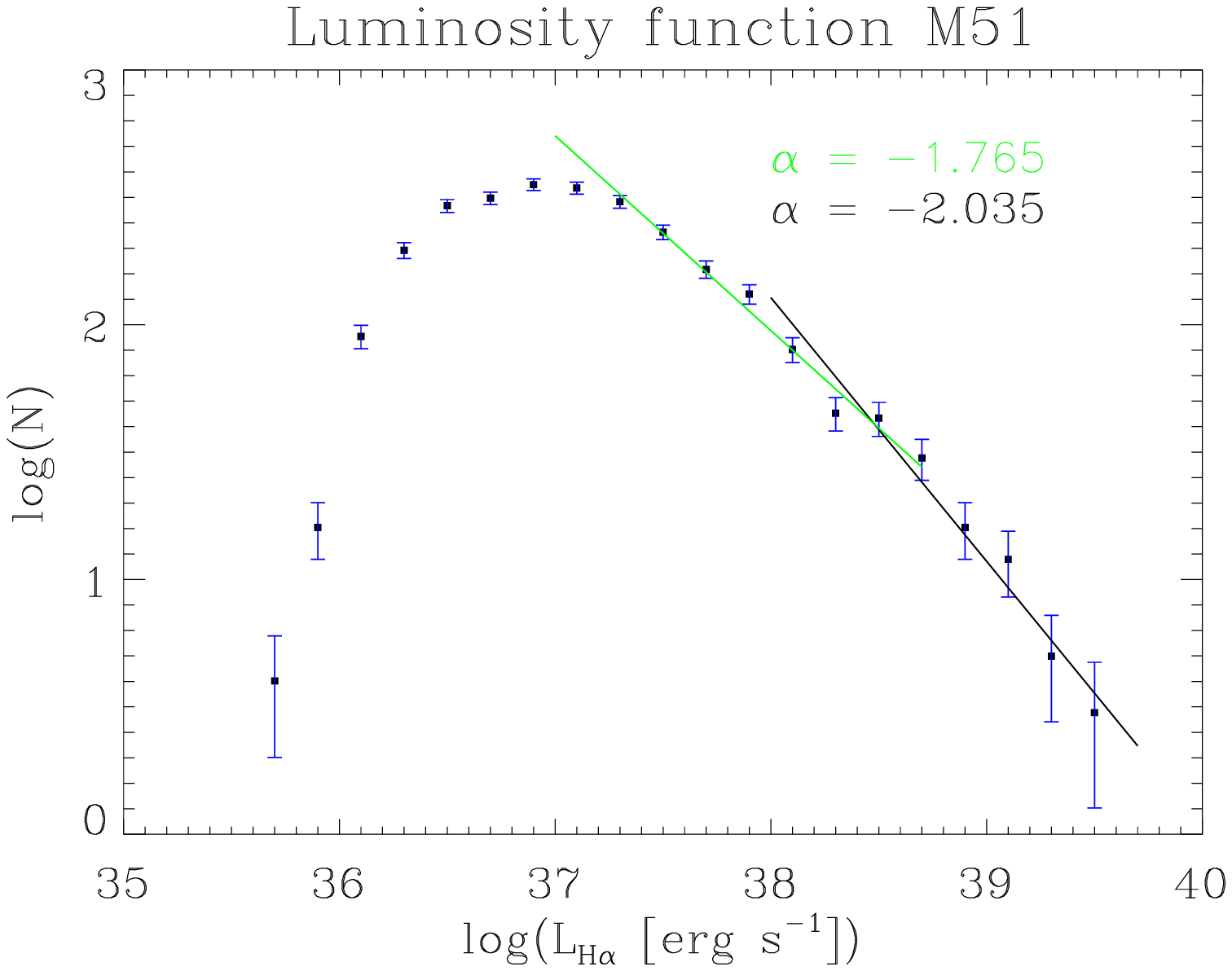}
   \caption{Luminosity function of the \hii{} regions of M51, as in Fig. \ref{F:lum_func}, where the falling portion at high luminosity has been represented by two linear fits, with a break point at luminosity 38.51 ($\pm{}$0.10) dex.} \label{F:lum_func_a}
\end{minipage}
\end{figure*} 

\begin{figure*}
\centering
\begin{minipage}{5.0in}
   \plotone{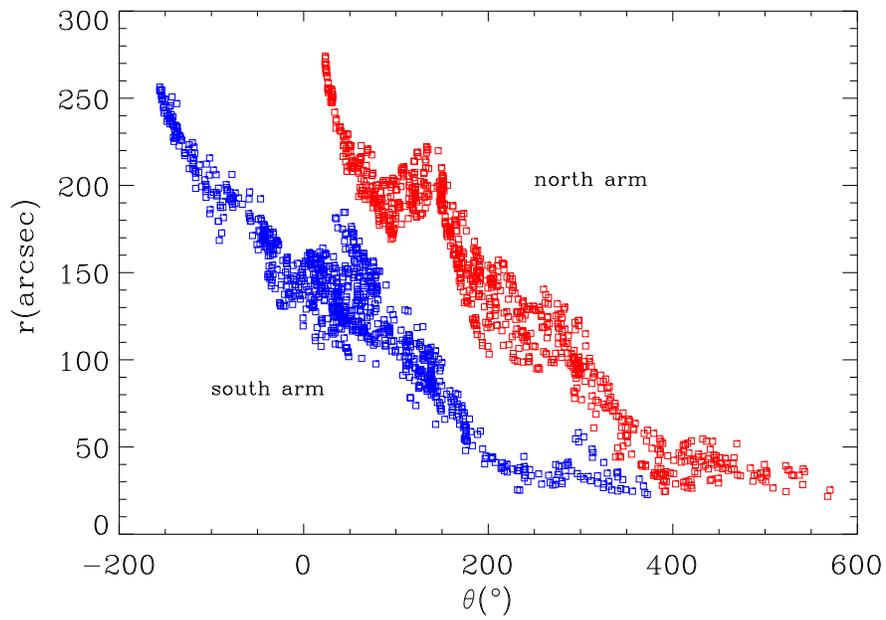}
   \caption{Distribution of the \hii{} regions of M51 in polar coordinates ($r$, $\theta$) showing the two arms of the galaxy. The coordinates were deprojected using  values of 42\degree~ for the inclination, and 26\degree~ for the position angle of the major axis, respectively. The \hii{} regions in the central 1 kpc zone were omitted from the plot.} \label{F:lum_func1}
\end{minipage}
\end{figure*} 

\begin{figure*}
\centering
\begin{minipage}{5.0in}
   \plotone{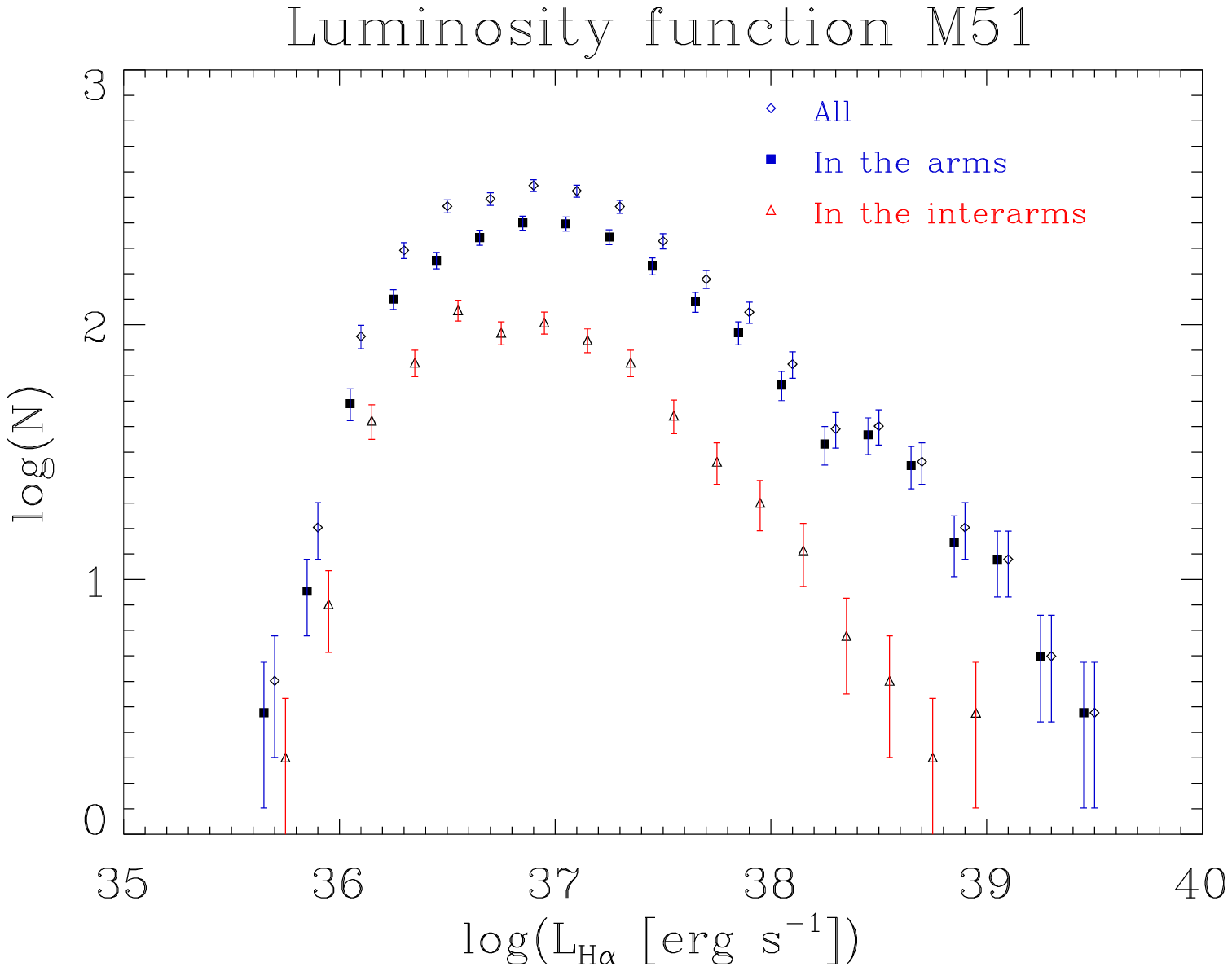}
   \caption{Luminosity function of the \hii{} regions of M51, including 
separately the functions for the arms and for the interarm region. For clarity, 
the function values for the arms are artificially displaced 0.05 dex to the 
left, and the values for the interarm are artificially displaced 0.05 dex to the 
right.} \label{F:lum_func2}
\end{minipage}
\end{figure*} 

We note that the population of the interarm zones is lacking in \hii{} regions in the high luminosity range. There are several possible explanations for this. 
One of them is that the mean age of the regions in the interarm zones is greater than the mean age of the regions in the arms, but this scenario appears improbable, because the only way we could envisage this is that the interarm \hii{} regions are the result of the migration of regions formed within the arms, so that the greater distances from the formation sites imply greater ages. However the numbers do not work out well here, since a characteristic lifetime for Lyman continuum emission in regions in the upper luminosity range is of order 3$\times$10$^6$ years, while the migration timescales implied are of order 3$\times$10$^7$ years. 
Another possibility is that in the arms we are seeing crowded regions so that the larger regions represent the overlapping of more than one region, a phenomenon which is infrequent in the interarm zones. However the evidence from \citet{relano05}, who obtained two dimensional velocity mapping of complete \hii{} region populations in galaxies using Fabry-Perot interferometry, and were thus able to separate the contributions of separate \hii{} regions subsumed into a major region, shows that in general the luminosity contribution of the smaller regions does not attain 30\% of the total integrated luminosity of the composite region. This is much less  than the order of magnitude difference between the most luminous arm regions and the most luminous interarm regions. 
So, the most probable reason for the difference is that the upper end of the main sequence in OB stars is more highly populated in the arms, because the placental cloud masses there are higher. But this would be an interesting point to pursue using observations of galaxies more local than M51.

\begin{figure*}
\centering
\begin{minipage}{5.0in}
   \plotone{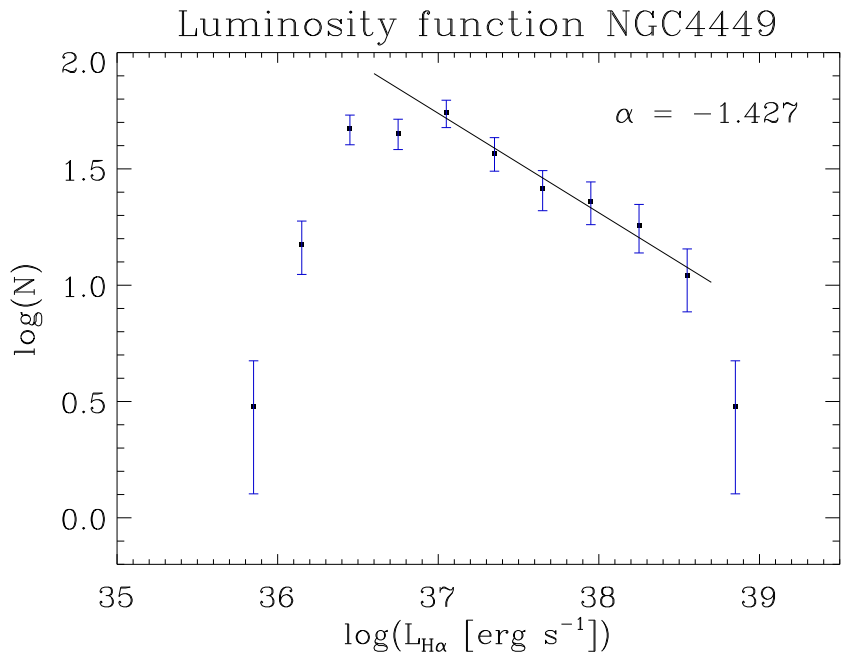}
   \caption{Luminosity function of the \hii{} regions of NGC 4449. We plotted 
the number $N$ of observed regions with luminosity between $L$ and $\Delta{L}$ 
(erg s$^{-1}$ on a logarithmic scale), with  $\Delta{L}$ = 0.3 dex.}  \label{F:lum_func_n4449}
\end{minipage}
\end{figure*}

In Fig. \ref{F:lum_func_n4449}
we show the luminosity function of the \hii{} regions measured in NGC 4449. The technique used was the same as that for M51 and needs little further description here. We can see that within the range of luminosity from $\log(L)$ = 37.0 to  $\log(L)$ = 38.5 dex (original units in erg s$^{-1}$) a power law fits the function quite well. The slope, $\alpha$, of this function has the value -1.43, which is slightly less negative that the mean value given by \citet{kennicutt89} for irregular galaxies, of -1.7, but differs by almost 0.5 from the values for NGC 4449 given by \citet{fuentes-masip00a}, and by \citet{valdez02}. Fuentes-Massip et al. acknowledge that their result, based on the analysis of only 44 \hii{} regions, is not statistically firm, and we consider that Valdez-Guti\'errez et al. were measuring blended regions (their data were from ground based observations) because their fit is to regions with luminosities greater than 38.6 dex, but we find only one region with a higher luminosity than this. In Table \ref{T:CD1} %
we show how varying the bin size of the luminosity function changes the value of $\alpha$, where we can see that changing the bin size over the range indicated does not have a major effect on the value of $\alpha$, and we finally selected the size which gave us the largest coefficient of determination, i.e. 0.3 dex in $\log(L)$. The integrated luminosity of the \hii{} regions measured in NCC4449 is 1.45$\times$10$^{40}$ erg s$^{-1}$, which is only 46\% of the total \ha{} luminosity of the whole galaxy. The integrated luminosity of the identified filamentary emission not counted as coming from \hii{} regions is only 3.74$\times$10$^{38}$ erg s$^{-1}$, which is only 1.2\% of the total. The remaining 53\% is emitted as diffuse emission, and our value agrees very well with that of \citet{kennicutt89} who give 52\%. Of this diffuse emission a little less than a half (44\%) is the underlying diffuse emission associated with the \hii{} regions, which we mentioned above. It is interesting to note here that in their detailed study of the diffuse emission in six local spirals, \citet{zurita04} found that a good ``rule of thumb'' is that approximately one half of the \ha{} emitted by a normal mid-type (Sb or Sc) spiral comes from the diffuse component, and the other half from the \hii{} regions.

\begin{table*} 
\begin{center}
\begin{minipage}{4.5in}
\begin{center}

\begin{tabular*}{4.0in}{@{\extracolsep{\fill}}c c  c}
\hline\hline
    \begin{tabular}{c} Bin \\
                           (dex)
      \end{tabular}  &   $\alpha$  &       
      \begin{tabular}{c} Coefficient of \\
                           determination
      \end{tabular}  \\
\hline\hline
             0.20   &    -1.49   &     0.736   \\
             0.22   &    -1.35   &     0.943   \\
             0.24   &    -1.39   &     0.928   \\
             0.26   &    -1.36   &     0.759   \\
             0.28   &    -1.54   &     0.871   \\
           {\bf 0.30}   &    {\bf -1.43}   &   {\bf  0.956}   \\
            0.32   &     -1.40   &     0.850   \\
            0.34   &     -1.46   &     0.906   \\
\hline
\end{tabular*} \\
\end{center}
\caption{\label{T:CD1} We show here the value of the exponent $\alpha$ for the power law adjusted to the luminosity function of the \hii{} regions of NGC 4449 as a function of the bin used. We also show the value of the coefficient of determination. The coefficient of determination has been calculated here simply as $R^2$, the square of the correlation coefficient between the actual values and the adjusted function. The values vary from 0 to 1.}
\end{minipage}
\end{center}
\end{table*}

\subsection{ The break in the luminosity function}

\begin{figure*}
\centering
\begin{minipage}{5.0in}
   \plotone{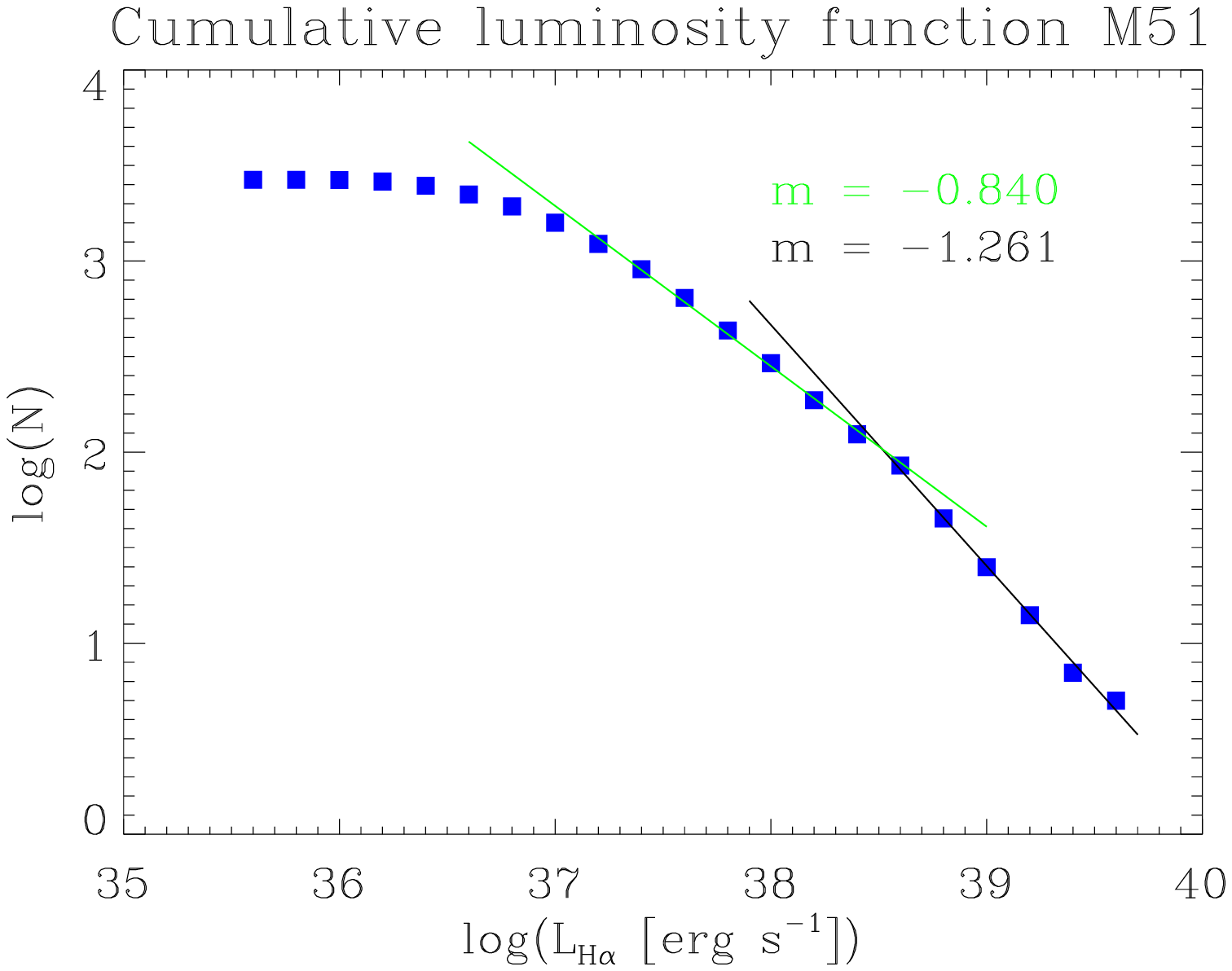}
   \caption{Cumulative luminosity function (the number of regions with luminosity greater than the value on the abscissa),
of the \hii{} regions of M51, showing clearly the break point at 38.5 dex.} \label{F:cumulative}
\end{minipage}
\end{figure*}

We note in Fig. \ref{F:lum_func_a} 
that the falling portion luminosity function of M51 at high luminosity could be well represented by a double linear fit, with a break point at luminosity 38.48 ($\pm{}$0.11) dex. 
The single fit described above has a coefficient of determination $R^2$ of 0.987, while the double fit gives a coefficient of determination $R^2$ of 0.984\footnote{The coefficient of determination for the double fit has been calculated taking the double fit as part of a single curve made by the union of the two segments.}, which does not indicate that the double line fit is superior. 
The break is, however, defined more cleanly in the cumulative luminosity function, which we show here in Fig. \ref{F:cumulative}.
Breaks of this kind were first noted by \citet{kennicutt89} who gave an estimated value for the break luminosity at $L$(\ha{}) = 38.7. They suggested, as did \citet{rand92} in his article on M51, and \citet{beckman00} that any galaxy would show the break, provided that it contained a sufficient number of \hii{} regions of high luminosity. The latter authors termed the luminosity at the break the ``Str\"omgren Luminosity''. \citet{rand92} found a value for this luminosity of $L$(\ha{}) = 38.6 dex and \citet{bradley06} in a study where they produced a summed luminosity function for almost 18,000 \hii{} regions in 57 galaxies derived exactly the same value for the luminosity of the break point.

\begin{figure*}
\centering
\begin{minipage}{5.0in}
   \plotone{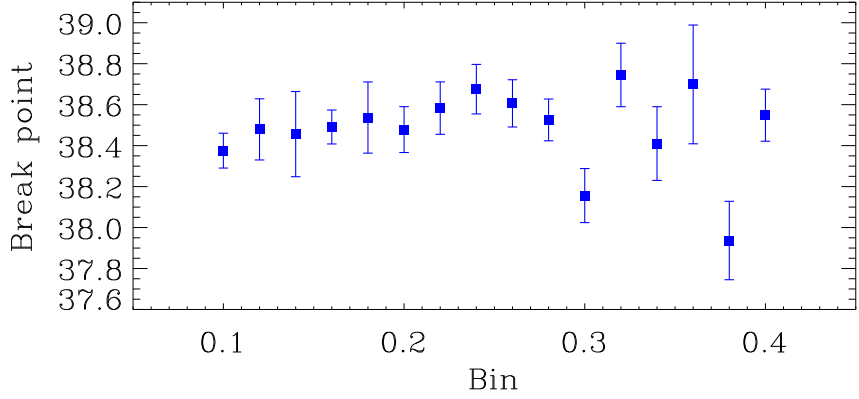}
   \caption{Plot of the derived values of the break luminosity against bin size. 
   We can see that the larger values of the bin size the larger dispersion in the 
   measured values.} \label{F:breaks}
\end{minipage}
\end{figure*}

\begin{figure*}
\centering
\begin{minipage}{5.0in}
   \plotone{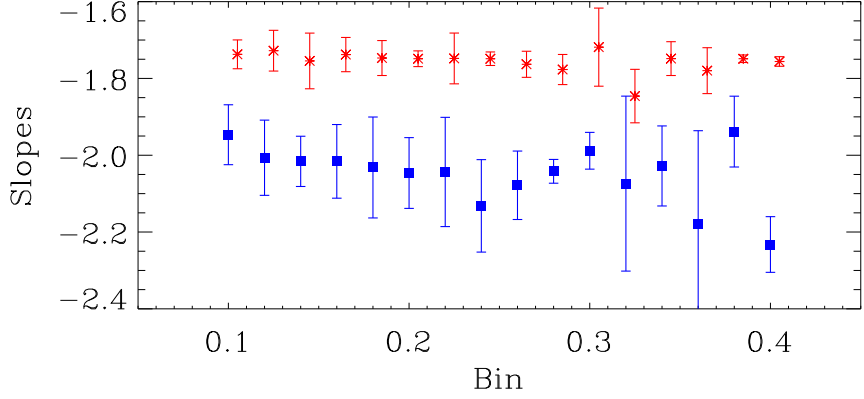}
   \caption{Plot of the values for the exponents of the two power laws fitted to the luminosity function varying the bin value from 0.1 to 0.4 dex. The points in the lower set correspond to the straight line fitted to the higher luminosity values (between 38.3 and 39.6 dex) while those in the upper set correspond to the values of intermediate luminosity (between 37.2 and 38.5 dex). The mean values of the exponents are -2.036 $\pm$ 0.048 and -1.749 $\pm$ 0.014.}
  \label{F:breaks_slopes}
\end{minipage}
\end{figure*}

In order to see to what extent the value of the break point is affected by the bin size, we plotted the derived values against bin size, as shown in Fig. \ref{F:breaks}
and as complementary information the slopes of the shallower and steeper parts of the gradient, also against bin size, in Fig. \ref{F:breaks_slopes}.
We can see that there is considerable stability in the value of the break luminosity for bins up to 0.28 dex in width, and a similar result holds for the slopes. Taking the values for the break luminosity in the range of bin width from 0.1 dex to 0.28 dex we find a value for the ``Str\"omgren Luminosity'' of 38.52 ($\pm{}$0.14) dex. Comparing our luminosity function with the very carefully measured ground based function of \citet{rand92}, we can see that the principal difference is that our sample is complete down to lower luminosities, and that we detect a larger number of small regions due to the superior resolution in the ACS images, all of which is as expected. Our study covers a slightly different fraction of the galaxy to that of Rand (the ACS mosaic does not include the southernmost portion of the galaxy, but given the improved resolution of our data we did not need to cut out such a large radial portion of the galaxy near the center to take care of crowding), which is sufficient reason for the small discrepancy in the value of the break luminosity, though this is within mutual error limits.

For NGC 4449 it is not possible to make a meaningful determination of a break luminosity, because of the paucity of high luminosity \hii{} regions, in an object which is much smaller and less massive than M51 (there are ten times fewer regions in NGC 4449). We can see in Fig. 6 that above $\log(L)$ = 38.52, the value of our break luminosity in M51, there is a sharp fall-off in the \hii{} region population of NGC 4449 too, although the numbers are too small to merit a statistical treatment.

\subsection{The diameter distribution for M51}

\begin{figure*}
\centering
\begin{minipage}{5.0in}
   \plotone{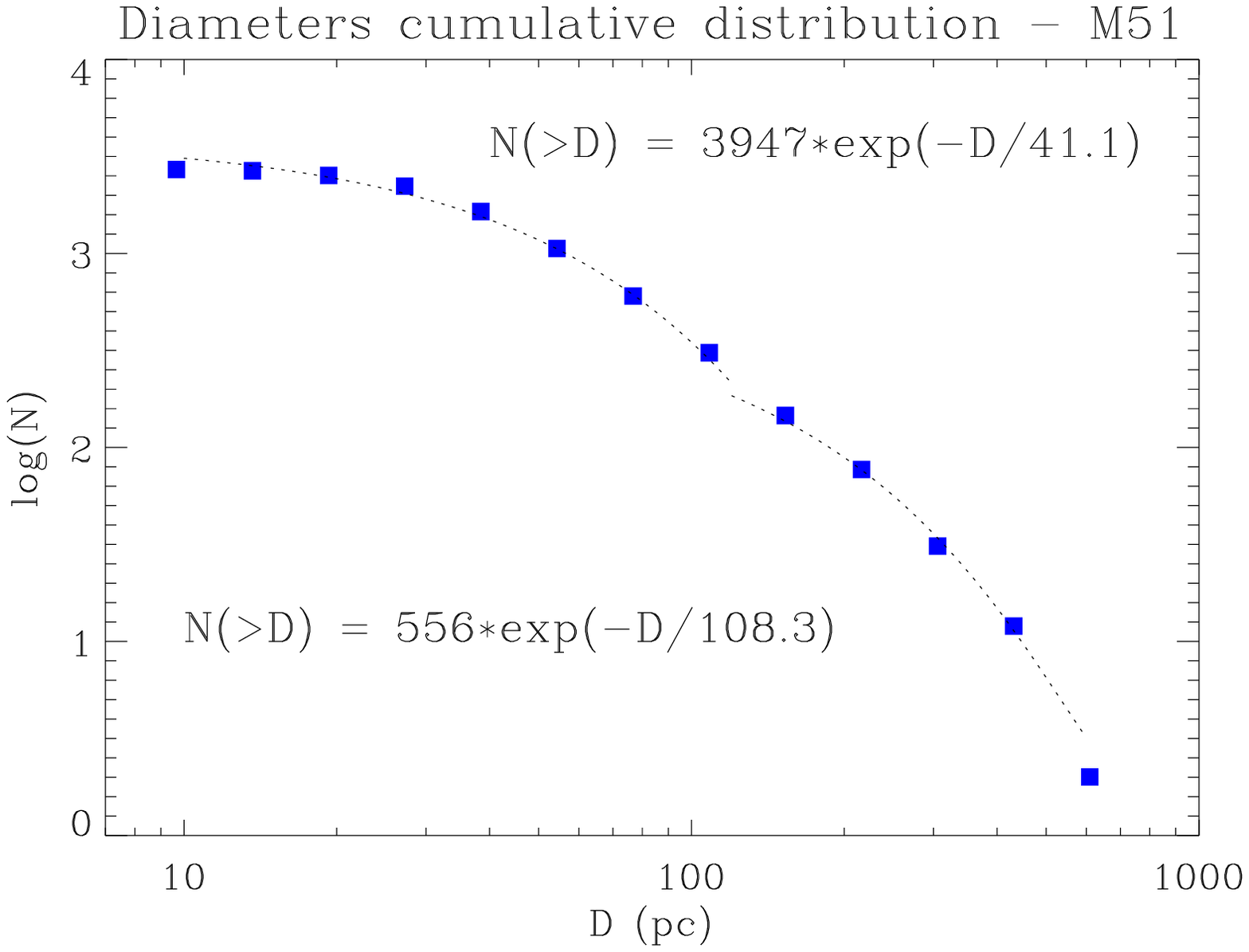}
   \caption{Cumulative distribution of the equivalent diameters of the \hii{} regions of the galaxy M51. The number $N$ of observed regions with equivalent diameter $D$ parsecs or greater is plotted.} \label{F:cumulative_diameters}
\end{minipage}
\end{figure*}

\begin{figure*}
\centering
\begin{minipage}{5.0in}
   \plotone{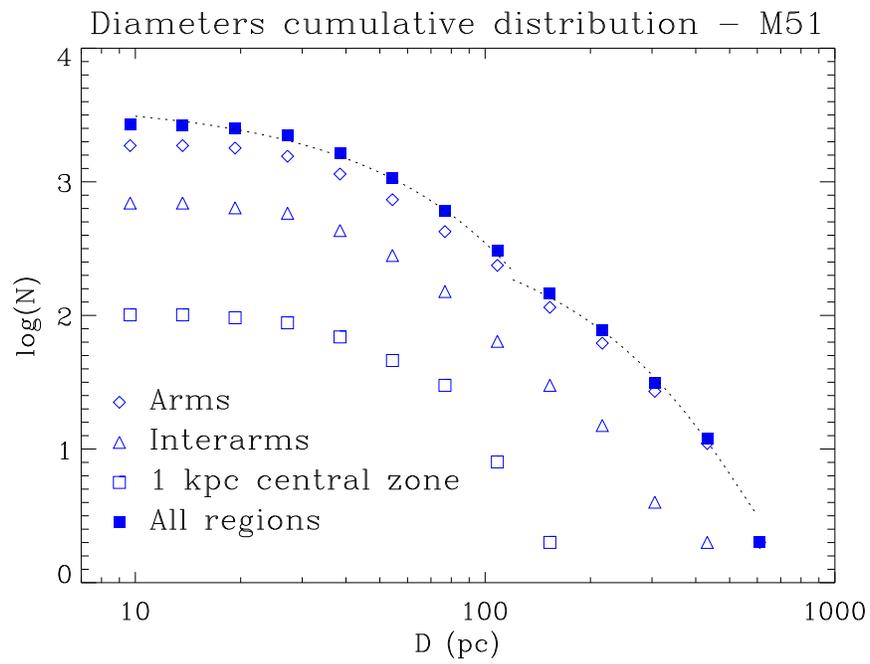}
   \caption{Cumulative distribution of the equivalent diameters of the \hii{} regions of the galaxy M51, as in Fig. \ref{F:cumulative_diameters}, plotting separately the regions in the arms, interarms, and the 1 kpc central zone.} \label{F:cumulative_diameters_separated}
\end{minipage}
\end{figure*}

In Fig. \ref{F:cumulative_diameters} we show the cumulative distribution of \hii{} region diameters for M51, defined as the sum of the numbers of regions with diameter greater than the plotted value. We have followed the functional form suggested first by van den Bergh (1981): $N = N_0 exp\left\{-(D/D_0)\right\}$ where $N$ is the integrated number of regions with diameters $D$ or greater, and $N_0$ and $D_0$ are constants which quantify the distribution. In M51 the range of diameters for the regions measured goes from 8.3 pc to 530 pc with a median value of 37.6 pc. If we fit a single function of this form to the points in Fig. \ref{F:cumulative_diameters} we find $D_0$ = 81.4 pc and $N_0$ = 2006. However, as shown in the figure, we find a better fit by combining two curves of the same form, and in that case, for the curve which fits best in the range $D <$ 120 pc the parameters found  are $D_0$ = 41.1 pc and $N_0$ = 3950 while for  $D >$ 120 pc the corresponding parameter values are $D_0$ = 108.3 pc while $N_0$ = 556. This graph suggests the possible presence of two population regimes. Following our discussion of the break in the luminosity function, we could envisage the break point between the curves as possibly marking the difference between \hii{} regions which are genuinely single, being ionized by a single OB cluster (the range of diameters approaching 120 pc corresponds to regions which are far too bright and large to be produced by even the most massive single star) and those which are compound, containing gas ionized by more than one cluster. Another possible explanation is that the high luminosity curve represents exclusively the \hii{} regions in the arms, while the lower luminosity curve represents the arm population with the addition of the regions in the interarm population. However, seeing Fig. \ref{F:cumulative_diameters_separated}, where we plot the cumululative distribution of \hii{} region diameters separating the regions in the arms, in the interarms and in the 1 kpc central zone, we can conclude that this is unlikely. The fact that the break feature appears in the interarm regions as well as in the arm regions seems to low the possibility that exist a population in the arm and another in the interarm. 

\clearpage

\section{Discussion}\label{sec:discussion}

Among the more interesting results presented in this article is the dependence of the \ha{} luminosity of the \hii{} regions, in both galaxies studied, on the square of the radius of the region, at least to zero order. This result is in conflict with the classical theory of the Str\"omgren sphere or similar models, which predict that for regions of constant gas density the luminosity would be proportional to the cube of the radius, and in this context the simplest way to account for the result is to assume that the mean density of an \hii{} region is inversely proportional to the square root of its luminosity. In fact an equivalent result was derived in a previous article, \citet{letter2010} using the mean electron density of a region as a measure of the mean density of the region. This global interpretation is strengthened by the measured dependence of the mean electron density of the \hii{} regions on galactocentric radius  \citep{letter2010}, which is a falling exponential with scale length equivalent to that of the neutral gas component. This demonstrates that, globally, the \hii{} region population is in pressure equilibrium with its local gas column in the disk, so that the electron density is a useful indicator of the overall mean density of an \hii{} region. However there is good evidence~\citep{oey97,relano02} that a significant fraction of the Lyman continuum photons produced by the OB stars within  the most luminous \hii{} regions escape into the surrounding diffuse gas, an inference which is at least partially supported by the fraction of \ha{} emitted by the diffuse ISM in NGC 4449 in the present article. This implies that a simple model for an \hii{} region with uniform density is not satisfactory, as is in any case known from the difference between the local electron density, measured from emission line ratios, and the global mean electron density, measured using the \ha{} luminosity and the radius. The factor of an order of magnitude between the two densities is classically explained \citep{osterbrock59} in terms of clumpy structure, with high density clumps producing the majority of the emission.

The fractional volume occupied by the clumps, the ``filling factor'' is routinely used in calculations of radiation balance and abundance determination in \hii{} regions. Any realistic model to explain the luminosity volume relation must take this inhomogeneous structure into account. 

A simple physical scheme would work as follows. If, at a given galactocentric radius the \hii{} regions are in quasi-pressure equilibrium, but the mean density varies as $1/R^{1/2}$ this implies that the mean temperature varies as $R^{1/2}$, which means that for a given luminosity $L$ the temperature varies as $L^{1/4}$. The more luminous the region the more it is heated and the higher its internal gas temperature, though for a factor of 10 rise in luminosity this amounts to a factor of only 1.78 in kinetic temperature of the gas. As a consequence of this the more luminous regions are more tenuous, and the mean spacing between clumps will grow, and the low interclump gas density will fall, which leads to an increase in photon escape from the region. It may well be possible to quantify this scenario further, but we must defer this until the effects of dust extinction on all of the radiative processes involved can be quantified, which is beyond the scope of the present article.


\section{Conclusions}\label{sec:conclusion}

We have used observations of two low redshift galaxies, M51 and NGC 4449, with the ACS on HST to quantify some of the principal physical characteristics of their populations of \hii{} regions. For both galaxies we have produced catalogs of the luminosities in \ha{}, the projected areas, and center positions of the \hii{} regions. For M51 the catalog contains 2659 regions, as the resolution and image quality of HST have allowed us to make a significant advance over previous ground based studies. Our catalog goes down 0.6 dex in luminosity compared to the detailed ground based study by \citet{rand92}, and the number of regions measured is greater by a factor 4. For NGC 4449 we analyzed the emission from 273 regions, which is not in fact a striking quantitative advance on ground-based catalogs \citep[see e.g.][]{sabbadin79} in this much less massive object with far fewer regions. From our databases we have derived the luminosity--volume (or luminosity--radius) relations for the regions of each galaxy, finding that in both cases the luminosity varies as the 2/3 power of the volume, (i.e. as the square of the radius) of the region. For a set of \hii{} regions with constant uniform density the relation between luminosity and volume would have been linear. We can account for our result if the \hii{} regions are globally in quasi pressure equilibrium with their surroundings, so that the mean electron density of a large and luminous region is less than that in a smaller less luminous region, because the mean temperature of the gas is larger in the former case. We have presented evidence for pressure equilibrium previously \citep{letter2010}, as the measured electron density in the \hii{} regions of our two measured galaxies varies with galactocentric distance with the same scale length as the neutral hydrogen, which would be the case for systems in pressure equilibrium. So there is a coherent qualitative explanation for the observed luminosity-volume relation, but a quantitative explanation for the simple relation found must await a more complete modelling exercise.

We also present luminosity functions for the \hii{} regions of both galaxies. Above the completeness limit, both functions show, to zero order approximation, a classical falling exponential. For M51 there is clear evidence for a break in the exponential, at a luminosity of $\log(L)$ = 38.51 (units in erg s$^{-1}$) with a steeper fall to higher luminosities confirming the previous finding by \citet{rand92}. The luminosity function for the interarm regions alone barely shows evidence for this break, but the number of highly luminous regions in this function is so small that it is not possible to state that the function is really distinct in form from the overall function for the galaxy, or from the function for the regions in the arms. For NGC 4449 there is also a steepening of the luminosity function above 38.5 dex, but here again the numbers of regions are too small to show whether this is a physical effect or merely a statistical effect. Explanations for the broken exponential, which has been reported previously in a number of papers, notably in \citet{kennicutt89}, have included the blending of \hii{} regions such that the largest regions absorb the smaller surrounding population \citep{pleuss00} and the increasing escape of ionizing photons from the larger regions \citep{beckman00}. The first hypothesis does yield a broken exponential but when developed in due detail predicts a shallower outer slope rather than the steeper slope observed. The evidence presented here that the more luminous a region the lower is its electron density does give some indirect support to the second hypothesis, but further work, notably incorporating the effects of dust extinction on the ionizing photon budget, will be needed to clarify the situation.

\section{Acknowledgements}\label{sec:acknowledgements}

This work was supported by grant P3/86 of the Instituto de Astrof\'isica de Canarias, and grant AYA-2007-67625-CO2-01 of the Spanish Ministry of Science and Innovation. LG thanks DGAPA-UNAM, M\'exico, for support provided through a PASPA fellowship, and the Instituto
de Astrof\'isica de Canarias for hospitality. 
This work is based on observations made with the NASA/ESA {\it Hubble Space Telescope},
obtained from the data archive at the Space Telescope Science Institute. STScI is
operated by the Association of Universities for Research in Astronomy Inc., under NASA
contract NAS5-26555.

\newpage{\pagestyle{empty}\clearpage}

\begin{deluxetable}{rrrrrrc}
\tablewidth{0pt}
\tablecaption{Brightest \hii{} Regions in M51\label{T:RegionsM51}}
\tablecolumns{7}
\tablehead{
  \colhead{Number} & \colhead{Rand} & \colhead{RA offset} &
  \colhead{DEC offset} & \multicolumn{2}{c}{Equivalent radius} &
    \colhead{Luminosity}  \\
  & & \colhead{(\arcsec{})}  & \colhead{(\arcsec{})} & \colhead{(\arcsec{})}  & \colhead{(pc)} &  
    \colhead{(10$^{38}$ erg s$^{-1}$)} \\
  \colhead{(1)} & \colhead{(2)} & \colhead{(3)} & \colhead{(4)} & \colhead{(5)} &
  \colhead{(6)} & \colhead{(7)} 
}

\startdata
       1  &        78  &       81.59  &     -104.35  &        6.57  &         267.6  &      52.341  \\ 
       2  &       319  &      -71.35  &      136.05  &        5.38  &         219.4  &      46.545  \\ 
       3  &        75  &       87.23  &      -79.65  &        3.45  &         140.7  &      39.884  \\ 
       4  &       308  &      -83.52  &       84.12  &        5.72  &         233.2  &      36.846  \\ 
       5  &       145  &      -98.55  &     -113.55  &        3.58  &         145.9  &      30.163  \\ 
       6  &        31  &        6.36  &       62.46  &        2.78  &         113.4  &      22.869  \\ 
       7  &       331  &      -29.18  &      139.15  &        3.25  &         132.7  &      18.857  \\ 
       8  &       308  &      -84.00  &       81.58  &        3.63  &         147.9  &      16.774  \\ 
       9  &       254  &      -41.54  &      -56.28  &        5.40  &         220.0  &      16.340  \\ 
      10  &       403  &       79.07  &      108.40  &        5.56  &         226.6  &      15.599  \\ 
      11  &       128  &      -60.50  &     -148.42  &        5.51  &         224.4  &      13.722  \\ 
      12  &       454  &      136.19  &     -182.12  &        3.77  &         153.8  &      13.377  \\ 
      13  &        30  &       -4.27  &       56.62  &        2.74  &         111.9  &      13.162  \\ 
      14  &       322  &      -60.63  &      147.18  &        3.12  &         127.0  &      12.273  \\ 
      15  &       141  &      -83.98  &     -133.35  &        2.26  &          92.1  &      11.585  \\ 
      16  &       146  &     -102.65  &     -105.72  &        2.60  &         106.1  &      10.990  \\ 
      17  &       523  &       23.68  &     -117.57  &        4.86  &         197.9  &      10.831  \\ 
      18  &       217  &     -115.41  &      231.44  &        3.04  &         124.0  &      10.727  \\ 
      19  &       269  &      -85.59  &       -2.84  &        3.95  &         161.0  &      10.599  \\ 
      20  &       320  &      -98.03  &      144.10  &        2.82  &         115.0  &      10.461  \\ 
      21  &       172  &     -146.05  &       -6.76  &        4.77  &         194.6  &       9.765  \\ 
      22  &        39  &       33.72  &       57.82  &        1.77  &          72.1  &       9.196  \\ 
      23  &       318  &      -62.65  &      128.18  &        2.62  &         106.7  &       8.732  \\ 
      24  &        22  &      -33.77  &       52.12  &        5.15  &         210.0  &       8.433  \\ 
      25  &       310  &      -78.80  &       95.88  &        2.52  &         102.6  &       7.720  \\ 
      26  &        44  &       63.25  &       50.69  &        4.01  &         163.3  &       7.652  \\ 
      27  &       400  &       57.42  &      117.56  &        2.52  &         102.7  &       7.446  \\ 
      28  &       218  &     -109.98  &      242.85  &        1.73  &          70.8  &       7.371  \\ 
      29  &       523  &       23.51  &     -118.68  &        3.39  &         138.4  &       7.190  \\ 
      30  &        21  &      -33.48  &       43.26  &        4.10  &         167.3  &       7.134  \\ 
      31  &       332  &      -29.62  &      129.50  &        2.36  &          96.4  &       6.722  \\ 
      32  &       148  &     -109.94  &     -121.36  &        2.12  &          86.7  &       6.721  \\ 
      33  &        69  &      109.73  &      -31.56  &        3.37  &         137.2  &       6.677  \\ 
      34  &       185  &     -148.21  &       45.66  &        4.67  &         190.4  &       6.519  \\ 
      35  &       284  &     -111.56  &       37.12  &        2.31  &          94.4  &       6.468  \\ 
      36  &       262  &      -71.13  &      -30.84  &        2.94  &         120.0  &       6.280  \\ 
      37  &         8  &      -34.39  &       12.55  &        2.40  &          97.8  &       6.239  \\ 
      38  &       190  &     -157.97  &       69.59  &        4.65  &         189.6  &       6.217  \\ 
      39  &       402  &       66.87  &      122.54  &        2.36  &          96.2  &       6.134  \\ 
      40  &        60  &       98.31  &       -5.66  &        3.27  &         133.5  &       5.974  \\ 
      41  &       309  &      -80.46  &       87.10  &        1.70  &          69.5  &       5.838  \\ 
      42  &       316  &      -71.06  &      117.38  &        2.45  &         100.1  &       5.818  \\ 
      43  &        79  &       60.41  &     -121.79  &        2.66  &         108.4  &       5.461  \\ 
      44  &       224  &       19.04  &       14.01  &        1.60  &          65.4  &       5.065  \\ 
      45  &         5  &      -31.63  &        2.07  &        1.27  &          51.8  &       4.987  \\ 
      46  &       161  &     -126.57  &      -68.98  &        4.40  &         179.3  &       4.874  \\ 
      47  &       342  &       23.13  &      123.09  &        2.22  &          90.6  &       4.856  \\ 
      48  &       313  &      -75.76  &      108.16  &        1.69  &          69.0  &       4.740  \\ 
      49  &   \nodata  &       12.10  &        0.41  &        1.93  &          79.0  &       4.663  \\ 
      50  &       194  &     -149.53  &       99.02  &        2.21  &          90.1  &       4.646 \\ 
\enddata

\tablecomments{Brightest \hii{} regions in the galaxy M51. The complete catalog is in the on-line version. 
(1) Consecutive number in the catalog; 
(2) Number of the \hii{} region in the \citet{rand92} catalog with which this region coincides;
(3) Offset in RA measured from the center of the galaxy;
(4) Offset in DEC measured from the center of the galaxy;
(5,6) Equivalent radius of the \hii{} region in arcsecs and parsecs, respectively, determined as explained in the text;
(7) Luminosity of the \hii{} region in units of 10$^{38}$ erg s$^{-1}$, assuming 8.4 Mpc as the distance to the galaxy M51.}

\end{deluxetable}

\begin{deluxetable}{rrrrrc}
\tablewidth{0pt}
\tablecaption{Brightest \hii{} Regions in NGC 4449\label{T:RegionsNGC}}
\tablecolumns{6}
\tablehead{
  \colhead{Number} & \colhead{RA offset} &
  \colhead{DEC offset} & \multicolumn{2}{c}{Equivalent radius} &
    \colhead{Luminosity}  \\
  & \colhead{(\arcsec{})}  & \colhead{(\arcsec{})} & \colhead{(\arcsec{})}  & \colhead{(pc)} &  
    \colhead{(10$^{38}$ erg s$^{-1}$)} \\
  \colhead{(1)} & \colhead{(2)} & \colhead{(3)} & \colhead{(4)} & \colhead{(5)} &
  \colhead{(6)} 
}

\startdata
       1  &     -29.92  &      93.33  &       3.7013  &        68.55  &        17.72981  \\ 
       2  &      20.29  &     -25.62  &       2.5352  &        46.95  &         7.14094  \\ 
       3  &      18.25  &     -19.99  &       1.3135  &        24.33  &         5.42239  \\ 
       4  &      20.73  &     -23.98  &       2.1504  &        39.83  &         4.57357  \\ 
       5  &      18.00  &     -17.10  &       1.6227  &        30.05  &         3.90805  \\ 
       6  &     -55.28  &      65.71  &       2.5092  &        46.47  &         3.67352  \\ 
       7  &      -0.96  &       0.63  &       1.3354  &        24.73  &         3.58257  \\ 
       8  &      20.45  &     -28.45  &       2.1279  &        39.41  &         3.57012  \\ 
       9  &     -21.56  &       5.94  &       1.7431  &        32.28  &         3.41079  \\ 
      10  &       8.74  &     -15.72  &       3.0259  &        56.04  &         3.39669  \\ 
      11  &     -15.41  &      84.50  &       2.8567  &        52.91  &         3.17414  \\ 
      12  &     -36.03  &      91.60  &       2.4026  &        44.50  &         2.52672  \\ 
      13  &     -40.81  &      92.86  &       1.4812  &        27.43  &         2.52531  \\ 
      14  &      21.91  &     -30.67  &       2.5068  &        46.43  &         2.45528  \\ 
      15  &     -16.30  &     -33.12  &       1.3986  &        25.90  &         2.23320  \\ 
      16  &       3.75  &      73.36  &       2.2074  &        40.88  &         2.18620  \\ 
      17  &      -2.41  &       4.28  &       1.5681  &        29.04  &         2.05437  \\ 
      18  &       6.74  &     -35.06  &       2.7877  &        51.63  &         2.05108  \\ 
      19  &     -44.29  &      92.27  &       2.1836  &        40.44  &         2.04145  \\ 
      20  &       2.60  &      -2.26  &       1.4348  &        26.57  &         1.97423  \\ 
      21  &     -70.90  &      52.01  &       1.9089  &        35.35  &         1.85603  \\ 
      22  &      -4.33  &       1.76  &       1.3710  &        25.39  &         1.78412  \\ 
      23  &     -27.74  &      35.02  &       1.8627  &        34.50  &         1.74252  \\ 
      24  &     -19.87  &     -37.70  &       1.4390  &        26.65  &         1.72020  \\ 
      25  &       7.16  &      69.40  &       1.5241  &        28.23  &         1.69270  \\ 
      26  &      13.34  &     -34.27  &       2.1042  &        38.97  &         1.49648  \\ 
      27  &      -7.07  &      17.36  &       1.4178  &        26.26  &         1.45747  \\ 
      28  &     -20.63  &      78.94  &       1.7068  &        31.61  &         1.38039  \\ 
      29  &     -20.60  &      -4.07  &       1.9507  &        36.13  &         1.37851  \\ 
      30  &      -3.19  &       0.28  &       0.8885  &        16.45  &         1.26688  \\ 
      31  &       8.34  &     -22.98  &       1.4585  &        27.01  &         1.25795  \\ 
      32  &     -18.29  &     -29.48  &       1.6403  &        30.38  &         1.24056  \\ 
      33  &     -14.16  &      10.10  &       1.0175  &        18.84  &         1.15385  \\ 
      34  &       2.90  &      -4.04  &       1.2518  &        23.18  &         1.08311  \\ 
      35  &      11.43  &     -12.54  &       1.1347  &        21.01  &         1.08194  \\ 
      36  &     -18.68  &      32.95  &       1.6325  &        30.23  &         1.07301  \\ 
      37  &     -71.37  &      56.60  &       1.6415  &        30.40  &         1.05303  \\ 
      38  &     -22.53  &     -29.95  &       1.8362  &        34.01  &         0.98065  \\ 
      39  &     -16.95  &     -10.27  &       1.7233  &        31.92  &         0.94258  \\ 
      40  &     -53.11  &      52.26  &       0.5578  &        10.33  &         0.92684  \\ 
      41  &       9.67  &     -19.41  &       1.5428  &        28.57  &         0.89394  \\ 
      42  &     -18.20  &      32.26  &       1.4062  &        26.04  &         0.84318  \\ 
      43  &     -40.49  &      63.99  &       1.4788  &        27.39  &         0.82203  \\ 
      44  &     -27.82  &       7.91  &       1.5327  &        28.39  &         0.82015  \\ 
      45  &      -5.96  &      21.15  &       1.3880  &        25.71  &         0.81475  \\ 
      46  &      47.48  &    -121.34  &       1.7087  &        31.65  &         0.80464  \\ 
      47  &      -2.83  &      52.26  &       1.3826  &        25.60  &         0.80276  \\ 
      48  &      14.05  &     -18.20  &       0.9262  &        17.15  &         0.72850  \\ 
      49  &     -24.49  &      91.10  &       0.7627  &        14.13  &         0.69442  \\ 
      50  &      18.33  &      63.85  &       1.3357  &        24.74  &         0.68432  \\ 
\enddata

\tablecomments{Brightest \hii{} regions in the galaxy NGC 4449. The complete catalog is in the on-line version. 
(1) Consecutive number in the catalog; 
(2) Offset in RA measured from the center of the galaxy;
(3) Offset in DEC measured from the center of the galaxy;
(4,5) Equivalent radius of the \hii{} region in arcsecs and parsecs, respectively, determined as explained in the text;
(6) Luminosity of the \hii{} region in units of 10$^{38}$ erg s$^{-1}$, assuming 3.82 Mpc as the distance to the galaxy NGC 4449.}

\end{deluxetable}


\section[]{On-line material}
The on-line material will soon be available on the AJ website. 

%
%
%
%
%
%
%
%


\begin{thebibliography}{}

\bibitem[Alonso-Herrero \& Knapen(2001)]{alonso01} Alonso-Herrero, A. \& Knapen J. H. 2001, AJ, 122, 1350

\bibitem[Annibali et al.(2008)]{annibali08} Annibali, F., Aloisi, A., Mack, J., et al. 2008, AJ, 135, 1900

\bibitem[Beckman et al.(2000)]{beckman00} Beckman, J. E., Rozas, M., Zurita, A., Watson, R. A., \& Knapen J. H. 2000, AJ, 119, 2728

\bibitem[Bersier et al.(1994)]{bersier94} Bersier, D., Belcha, A., Golay, M., \& Martinet, L. 1994, A\&A, 286, 37

\bibitem[B\"oker et al.(1999)]{boker99} B\"oker. T., Calzetti, D.,  Sparks, W., et al. 1999, ApJS, 124, 95

\bibitem[B\"oker et al.(2001)]{boker01} B\"oker, T., van der Marel, R. P., Mazzuca, L., Rix, H.-W., Rudnick, G., Ho, L. C., \& Shields, J. C. 2001, AJ, 121, 1473

\bibitem[Bradley et al.(2006)]{bradley06} Bradley, T. R., Knapen, J. H., Beckman, J. E., \& Folkes, S. L. 2006, A\&A, 459, 13L

\bibitem[Bresolin et al.(2004)]{bresolin04} Bresolin, F., Garnett, D. R., \& Kennicutt, R. C. 2004, ApJ, 615, 228

\bibitem[Buckalew \& Kobulnicky(2006)]{buckalew06} Buckalew, B. A. \& Kobulnicky, H. A. 2006, AJ, 132, 1061

\bibitem[Cardwell et al.(2000)]{cardwell00} Cardwell, A., Beckman, J.E., Magrini, L., \& Zurita, A. 2000, Proceedings 232 WE-Heraeus Seminar ``The Interstellar Mediuim in M3, and M33'' (Eds. Elly M. Berkhuijsen, Rainer Beck, and Rene A. M. Walterbos), Shaker, Aachen, p. 115

\bibitem[Feinstein(1997)]{feinstein97} Feinstein, C. 1997, ApJS, 112, 29

\bibitem[Feldmeier et al.(1997)]{feldmeier97} Feldmeier J. J., Ciardullo R., \& Jacoby G. H. 1997, ApJ, 479, 231

\bibitem[Fuentes-Masip et al.(2000)]{fuentes-masip00a}  Fuentes-Masip, O., Casta\~neda, H. O., \& Mu\~noz-Tu\~n\'on, C. 2000, AJ, 119, 2166

\bibitem[Gonz\'alez-Delgado \& P\'erez(1997)]{gonzalez97} Gonz\'alez-Delgado, R. M. \& P\'erez, E. 1997, ApJS, 108, 199

\bibitem[Guti\'errez \& Beckman(2010)] {letter2010} Guti\'errez, L. \& Beckman, J. E., 2010, ApJL, 710, 44

\bibitem[Hack et al(2000)]{hack00} Hack, W.J. \& Greenfield, P. 2000, {\it ASP Conf. Ser., Vol. 216, Astronomical Data Analysis Software and Systems IX}, eds. N. Manset, C. Veillet, \& D. Crabtree (San Francisco: ASP), 433

\bibitem[Hakobyan et al.(2008)]{hakobyan08} Hakobyan, A. A., Petrosian, A. R., Yeghazaryan, A. A., \& Boulesteix, J. 2008, Ap, 50, 426 

\bibitem[Hill et al.(1997)]{hill97} Hill, J. K., Waller, W. H., Cornett, R. H., et al. 1997, ApJ, 477, 673

\bibitem[Hodge, Lee, \& Kennicutt(1989)]{hodge89} Hodge, P., Lee, M. G., \& Kennicutt, R. C. 1989, PASP, 101, 32 

\bibitem[Karachentsev et al.(2003)]{karachentsev03} Karachentsev, I. D., Sharina, M. E., Dolphin, A. E., et al. 2003, A\&A, 398, 467

\bibitem[Kennicutt(1992)]{kennicutt92} Kennicutt, R. C. 1992, ApJ, 388, 310

\bibitem[Kennicutt \& Hodge(1980)]{kennicutt80} Kennicutt, R. C. \& Hodge, P. W. 1980, ApJ, 241, 573 

\bibitem[Kennicutt et al.(1989)]{kennicutt89} Kennicutt, R. C., Edgar, B. K., \& Hodge, P. W. 1989, ApJ, 337, 761

\bibitem[Knapen et al.(1993)]{knapen93} Knapen, J. H., Arnth-Jensen, N., Cepa, J., \& Beckman, J. E. 1993, AJ, 106, 56

\bibitem[Knapen et al.(2004)]{knapen04} Knapen, J. H., Stedman, S., Bramich, D. M., Folkes, S. L., \& Bradley, T. R. 2004, A\&A, 426, 1135

\bibitem[Koekemoer et al.(2002)]{koek02} {Koekemoer, A. M., Fruchter, A. S., Hook, R. N., \& 
     Hack, W. 2002, HST Calibration Workshop, Ed. S. Arribas, A. M. Koekemoer, B. Whitmore 
     (STScI: Baltimore), 337. In http://www.stsci.edu/}

\bibitem[Lisenfeld \& Ferrara(1998)]{lisenfeld98} Lisenfeld, U., \& Ferrara, A. 1998, ApJ, 496, 145

\bibitem[MacKenty et al.(2000)]{mackenty00} MacKenty, J. W., Ma\'iz-Apell\'aniz, J., Pickens, C. E., Norman, C. A., \& Walborn, N. R. 2000, AJ, 120, 3007

\bibitem[Mutchler et al.(2005)]{mutchler05} {Mutchler, M., Beckwith, S.V.W., Bond, H. E., 
     Christian, C., Frattare, L., Hamilton, F.,
     Hamilton, M., Levay, Z., Noll, K., \& Royle, T.
     2005,  ``Hubble Space Telescope multi-color ACS mosaic
     of M51, the Whirlpool Galaxy'',
     Bulletin of the American Astronomical Society, Vol.37, No.2. 
     http://www.aas.org/publications/baas/v37n2/ aas206/339.htm}

\bibitem[Oey \& Kennicutt(1997)]{oey97} Oey, M. S. \& Kennicutt, R. C. 1997, MNRAS, 291, 827

\bibitem[Osterbrock(1989)]{osterbrock89} {Osterbrock, D. 1989, in {\it Astrophysics 
     of Gaseous Nebulae and Active Galactic Nuclei}, Mill Valley: University Science Books}

\bibitem[Osterbrock \& Flather(1959)]{osterbrock59} Osterbrock, D. \& Flather, E. 1959, ApJ, 129, 260
	
\bibitem[Pleuss, Heller, \& Fricke(2000)]{pleuss00} Pleuss, P. O., Heller, C. H., \& Fricke, K. J. 2000, A\&A, 361, 913 

\bibitem[Rand(1992)]{rand92}           Rand, R. J. 1992, AJ, 103, 815

\bibitem[Rela\~no et al.(2002)]{relano02} Rela\~no, M., Peimbert, M., \& Beckman, J. 2002, ApJ, 564, 704

\bibitem[Rela\~no et al.(2005)]{relano05} Rela\~no, M., Beckman, J. E., Zurita, A., Rozas, M., \& Giammanco, C. 2005, A\&A, 431, 235

\bibitem[Rozas et al.(1996)]{rozas96} Rozas, M., Beckman, J. E., \& Knapen, J. H. 1996, A\&A, 307, 735

\bibitem[Rozas et al.(1999)]{rozas99} Rozas, M., Zurita, A., Heller, C. H., \& Beckman, J. E. 1999, A\&AS, 135, 145

\bibitem[Rozas et al.(2000)]{rozas00} Rozas, M., Zurita, A., \& Beckman, J. E. 2000, A\&A, 354, 823
   
\bibitem[Sabbadin \& Bianchini(1979)]{sabbadin79} Sabbadin, F. \& Bianchini, A. 1979, PASP, 91, 280

\bibitem[Scoville et al.(2001)]{scoville01} Scoville, N. Z., Polletta, M., Ewald, S., Stolovy, S. R., Thompson, R., \& Rieke, M. 2001, AJ 122, 3017 %


\bibitem[Sirianni et al.(2005)]{sirianni05}  Sirianni, M., Jee, M. J., Ben\'itez, N., et al. 2005, PASP, 117, 1049

\bibitem[Str\"omgren(1939)]{stromgren39} Str\"omgren, B. 1939, ApJ, 89, 526

%

\bibitem[Thilker et al.(2002)]{thilker02} Thilker, D. A., Walterbos, R. A. M., Braun, R., \& Hoopes, C. G. 2002, AJ, 124, 3118


\bibitem[Tonry et al.(2001)]{tonry01} Tonry, J. L., Dressler, A., Blakeslee, J. P., et al. 2001, ApJ, 546, 681

\bibitem[Valdez-Guti\'errez et al.(2002)]{valdez02} Valdez-Guti\'errez, M., Rosado, M., Puerari, I., Georgiev, L, Borisova, J., \& Ambrocio-Cruz, P. 2002, AJ, 124, 3157

\bibitem[Youngblood \& Hunter(1999)]{youngblood99} Youngblood, A. J. \& Hunter, D. A. 1999, ApJ, 519, 55

\bibitem[Zurita et al. (2004)]{zurita04} Zurita, A., Rela\~no, M., Beckman, J. E., \& Knapen, J.H. 2004, A\&A, 413, 73
%
%
%

%
%
%
%
%
%

\end{thebibliography}
\end{document}